
\input harvmac.tex
\input epsf

\def\figin{\epsfcheck\figin}\def\figins{\epsfcheck\figins}
\def\epsfcheck{\ifx\epsfbox\UnDeFiNeD
\message{(NO epsf.tex, FIGURES WILL BE IGNORED)}
\gdef\figin##1{\vskip2in}\gdef\figins##1{\hskip.5in}
\else\message{(FIGURES WILL BE INCLUDED)}%
\gdef\figin##1{##1}\gdef\figins##1{##1}\fi}
\def\DefWarn#1{}
\def\figinsert{\goodbreak\midinsert}
\def\ifig#1#2#3{\DefWarn#1\xdef#1{fig.~\the\figno}
\writedef{#1\leftbracket fig.\noexpand~\the\figno}%
\figinsert\figin{\centerline{#3}}\medskip\centerline{\vbox{\baselineskip12pt
\advance\hsize by -1truein\noindent\footnotefont{\bf Fig.~\the\figno:} #2}}
\bigskip\endinsert\global\advance\figno by1}

\def\Title#1#2{\rightline{#1}
\ifx\answ\bigans\nopagenumbers\pageno0\vskip0.5in%
\baselineskip 15pt plus 1pt minus 1pt
\else
\pageno1\vskip.25in\fi \centerline{\titlefont #2}\vskip .3in}

\ifx\answ\bigans\def\tcbreak#1{}\else\def\tcbreak#1{\cr&{#1}}\fi

\font\cmss=cmss10 \font\cmsss=cmss10 at 7pt
\def\IZ{\relax\ifmmode\mathchoice
{\hbox{\cmss Z\kern-.4em Z}}{\hbox{\cmss Z\kern-.4em Z}}
{\lower.9pt\hbox{\cmsss Z\kern-.4em Z}}
{\lower1.2pt\hbox{\cmsss Z\kern-.4em Z}}\else{\cmss Z\kern-.4em Z}\fi}
\def\calm{{\cal M}}

\lref\REFmatrixmodels{
D. Gross and A. Migdal,  Nucl. Phys. {\bf B340} (1990) 333\semi
E. Br\'ezin and V. Kazakov, Phys. Lett. {\bf B236} (1990) 144\semi
M. Douglas and S. Shenker, Nucl. Phys. {\bf B335} (1990) 635.}

\lref\REFcequalsone{
D. Gross and N. Miljkovi\'c, Phys. Lett. {\bf B238} (1990) 217\semi 
E. Br\'ezin, V. Kazakov and  A. Zamolodchikov, \hfill\break
Nucl. Phys. {\bf B338} (1990) 673.}

\lref\REFgrossreview{
D. Gross, {\sl The $c=1$ Matrix Models,} in: {\sl Two Dimensional
Quantum Gravity and Random Surfaces} (World Scientific, Singapore, 1992) p.143}

\lref\REFdavid{
F. David, {\sl A Scenario for the  $c>1$ Barrier in
Noncritical Bosonic Strings,} report SACLAY-SPHT-96-112, October 1996,
{\tt hep-th/9610037}.}

\lref\REFEKreduction{
T. Eguchi and  H. Kawai, Phys. Rev. Lett. {\bf 48} (1982) 1063\semi
A. Migdal, Phys. Rep. {\bf 102} (1983) 199.}
 
\lref\REFparisioriginal{
G. Parisi, Phys. Lett. {\bf B238} (1990) 213.}

\lref\REFkostov{
I. Kostov, Nucl. Phys. {\bf B376} (1992) 539.}

\lref\REFparisielse{
G. Parisi, Phys. Lett. {\bf B238} (1990) 209;
Europhys. Lett. {\bf 11} (1990) 595\semi
G. Parisi, G. Salina and A. Vladikas, Phys. Lett. {\bf B256} (1991) 397.}

\lref\REFgrossklebanov{D. Gross and I. Klebanov, 
Nucl. Phys. {\bf B344} (1990) 475\semi
Nucl. Phys. {\bf B354} (1991) 459.}

\lref\REFboulatovkazakov{
D. Boulatov and V. Kazakov, Int. J. Mod. Phys. {\bf A8} (1993) 809\semi
Nucl. Phys. Proc. Suppl. {\bf 25A} (1992) 38.} 

\lref\REFkosterlitzthouless{
V. Berezinskii, JETP {\bf 34} (1972)610\semi
J.M. Kosterlitz and D. Thouless, J. Phys. {\bf C6} (1973) 1181\semi
J. Villain, J. Phys. {\bf C36} (1975) 581.}

\lref\REFgrosswitten{
D. Gross and E. Witten, Phys. Rev. {\bf D21} (1980) 446.}

\lref\REFdouglaskazakov{
M. Douglas and V. Kazakov, Phys. Lett. {\bf B319} (1993) 219.}

\lref\REFbipz{
E. Br\'ezin, C. Itzykson, G. Parisi and  J.B. Zuber,\hfill\break
Commun. Math. Phys. {\bf 59} (1978) 35.}

\lref\REFitzyksonzuber{
C. Itzykson and J.B. Zuber, J. Math. Phys. {\bf 21} (1980) 411.}

\lref\REFhopf{
A. Matytsin, Nucl. Phys. {\bf B411} (1994) 805.}

\lref\REFgrossinstantons{
D. Gross and A. Matytsin, 
Nucl. Phys. {\bf B429} (1994) 50\semi 
Nucl. Phys. {\bf B437} (1995) 541\semi
J. Minahan and A. Polychronakos,  Nucl. Phys. {\bf B422} (1994) 172.}

\lref\REFwynter{
V. Kazakov, M. Staudacher and T. Wynter,\hfill\break
Commun. Math. Phys. {\bf 177} (1996) 451\semi
Nucl. Phys. {\bf B471} (1996) 309.}

\lref\REFpenner{
Yu. Makeenko, Phys. Lett. {\bf B314} (1993) 197\semi
Int. J. Mod. Phys. {\bf A10} (1995) 2615\semi
L. Paniak and  N. Weiss, J. Math. Phys. {\bf 36} (1995) 2512.}

\lref\REFlittleknown{
D. Boulatov, Mod. Phys. Lett. {\bf A9} (1994) 1963\semi
A. Matytsin and A. Migdal, Int. J. Mod. Phys. {\bf A10} (1995) 421\semi
J.-M. Daul, {\sl $Q$-states Potts Model on a Random Planar Lattice,}
ENS report, November 1994, {\tt hep-th/9502014}.}

\lref\REFgrossinducedQCD{
D. Gross, Phys.Lett. {\bf B293} (1992) 181.}

\lref\REFkazakovmulticritical{
V. Kazakov, Mod. Phys. Lett. {\bf A4} (1989) 2125.}

\Title{\vbox{\baselineskip12pt\hbox{MIT-CTP-2589}
\hbox{ENSLAPP-A-627/96}}}
{\vbox{\centerline{Kosterlitz--Thouless Phase Transitions }\vskip0.15in
\centerline{on Discretized Random Surfaces }}}
\centerline{\vbox{\hsize3in\centerline{Andrei Matytsin}}}
{\it
\smallskip
\centerline{Center for Theoretical Physics}
\centerline{Laboratory for Nuclear Science}
\centerline{Massachusetts Institute of Technology}
\centerline{77 Massachusetts Avenue}
\centerline{Cambridge, MA 02139}}
\smallskip
\centerline{and}
\smallskip
\centerline{
\vbox{\hsize3in\centerline{Philippe Zaugg}}}
%
{\it
\smallskip
\centerline{Laboratoire de Physique Th\'eorique
ENSLAPP\footnote
{$^*$}{URA 14-36 du CNRS, associ\'ee \`a l'Ecole Normale Sup\'erieure de
Lyon et \`a l'Universit\'e de Savoie.}}
\centerline{LAPP, Chemin de Bellevue, BP 110}
\centerline{F-74941 Annecy-le-Vieux Cedex, France\footnote{}{e-mail:
{\tt matytsin@marie.mit.edu, zaugg@lapp.in2p3.fr}}
}}
\smallskip
\bigskip
\centerline{\bf Abstract}
\bigskip
\noindent
The large $N$ limit of a one-dimensional infinite chain 
of random matrices is investigated. It is found 
that in addition to the expected 
Kosterlitz--Thouless phase transition this model exhibits an infinite 
series of phase transitions at special values of the lattice spacing 
$\epsilon_{pq}
=\sin(\pi p/2q).$ An unusual property of these transitions is that they
are totally invisible in the double scaling limit. A method
which allows us to explore the transition regions analytically and to
determine certain critical exponents is developed. It is argued that 
phase transitions of this kind can be induced by the interaction of 
two-dimensional vortices with curvature defects of a fluctuating
random lattice.

\secno 0
\newsec{Introduction}
After a period of considerable progress in the study 
of two-dimensional quantum
gravity and noncritical strings \REFmatrixmodels\
several fundamental problems in the 
field remained open.

The most prominent among them is, perhaps, the problem of the $c=1$
barrier \REFcequalsone \REFgrossreview. 
Indeed, string theories with the central charge of the matter
$c<1$ are by now rather well understood.
On the contrary, very little is known about $c>1$ strings, but their 
physical properties are probably quite different \REFdavid.

The 
problematic nature of $c>1$ theories can be partly attributed to the
presence of a tachyon in their spectra. However, 
this is not the main obstacle in studying them.
Indeed, in the random matrix model approach to 
string theory the cause of the difficulties
appears to be more technical \REFgrossreview. 
In fact, it is perfectly possible 
to write down a matrix
model which in the large $N$ limit reproduces the genus expansion
of a $D$-dimensional noncritical string with $D>1$. 
What is lacking are the analytic tools necessary to explore such models.

The reason is that, compared to $c<1$ models, 
string theories with $c$ larger than one have  
significantly more degrees of freedom. In the continuum formulation these 
additional degrees of freedom correspond to the transverse modes of the string,
absent for $c<1$. 
The appearance of new modes is especially clear in the matrix model 
language.
The dynamical variables of matrix models are Hermitian $N\times N$ matrices,
each characterized by  $N^2$ independent matrix elements. 
If we diagonalize a Hermitian matrix,
representing it as $M=U \lambda U^{\dagger}$, these $N^2$ degrees of freedom
can be split into $N$ eigenvalues forming the diagonal matrix $\lambda$ and
$N^2-N$ ``angular variables'' encoded in the unitary matrix $U$. In $c<1$
matrix
models the angular degrees of freedom decouple (or can 
be easily integrated out),
leaving us with a theory of $N$ eigenvalues which contains all the information 
about the original string theory. 
For matrix models of $D>1$ strings this no longer occurs.
As a result, in more than one dimension 
the dynamics of the theory depends on
all of $N^2\gg N$ variables. 


Mathematically, 
when the angular variables do not decouple the partition function 
of a matrix model involves a nontrivial integral over unitary matrices~$U$.
Explicit evaluation of such integrals would be useful for applications
not only in matrix models but also in lattice gauge theories. In fact, the 
large $N$ limit of QCD can be interpreted, through the Eguchi--Kawai
reduction, as a certain integral over four 
unitary matrices \REFEKreduction. 
Thus even the large $N$ QCD can be viewed as a very 
complicated multimatrix model, where the unitary degrees of freedom play 
a crucial role.

In this paper we shall explore the effects of such degrees of freedom in 
the simplest model where these effects are nontrivial. 
The theory to be considered is a one-dimensional infinite chain of Hermitian
matrices $M_n$ defined by the partition function\foot{In this formula $V(M)$
is a polynomial potential such as, for example, $m^2 M^2/2+g M^3/3$ or
$m^2 M^2/2 + g M^4/4$.}
\eqn\chain{
{\cal Z}=\int\prod\limits_{n=-\infty}^{+\infty}\!\! dM_n\thinspace
\exp\biggl\{ -N\thinspace {\rm Tr}\sum_{n=-\infty}^{+\infty} 
\biggl[{(M_{n+1}-M_n)^2\over 2\epsilon} + \epsilon V(M_n)\biggr]
\biggr\}.}
This model is, in fact, quite interesting by itself 
\REFparisioriginal\REFgrossklebanov\REFkostov. It represents 
a one-dimensional string theory whose target space, instead of being 
continuous, consists of an infinite number of equidistant discrete points. 
Furthermore, as we shall discuss below, the large $N$ limit of this model
is related by a duality transformation to the two-dimensional 
$O(2)$ nonlinear sigma model coupled to quantum gravity or, equivalently, 
to one-dimensional bosonic string theory compactified
on a circle of radius $R=1/\epsilon$.

The multimatrix chain exhibits a phase transition induced precisely by the
unitary degrees of freedom discussed above 
\REFgrossklebanov\REFboulatovkazakov. Indeed,
in the limit of infinitely small lattice spacing $\epsilon\to 0$
our model describes the one-dimensional 
string theory with a continuous target space.
Such a theory has $c=1$. On the other hand, when the lattice spacing gets
very large the interaction of neighboring matrices 
${\rm Tr}(M_n-M_{n+1})^2/2\epsilon$ becomes negligible compared to
$\epsilon\thinspace{\rm Tr}V(M)$. Consequently, at $\epsilon=\infty$
the multimatrix chain
decouples into a product of infinitely many identical one-matrix models.
For cubic $V(M)$ these one-matrix models represent pure two-dimensional
quantum gravity, a system with $c=0$.  
Therefore, the central charge of string theory corresponding to \chain\
changes from one to zero as $\epsilon$ is increased. In fact, there is 
evidence that this change occurs sharply at a certain value of the lattice 
spacing $\epsilon=\epsilon_{\rm cr}$ where the model undergoes a phase 
transition. 

Of course, being one-dimensional the matrix chain is substantially simpler 
than any of the prospective $c>1$ models. However, we shall see that already 
in this simple theory the unitary variables produce new effects not encountered
at $c<1$.
%
%
%
Surprisingly, we find that the phase transition at 
$\epsilon=\epsilon_{\rm cr}$ is not the only critical point of \chain.
It turns out that in addition 
this model has an infinity of critical
points at $\epsilon=\epsilon_{\rm cr}\sin(\pi p/2q)$ labelled by
positive integers
$p$ and $q$, 
where certain observables, such as the
eigenvalue densities of matrices $M_n$, develop singularities. These
singularities are of universal nature (that is, they are the same for various
potentials $V(M)$ within a certain class) and evidence phase transitions.
It is quite unusual, however, that none of these transitions affects
the double scaling limit of the theory so that \chain\ has $c=1$ for any
$\epsilon<\epsilon_{\rm cr}.$

It is rather amusing that the additional critical points fill densely the
interval $[0, \epsilon_{\rm cr}].$ This shows that once the unitary
degrees of freedom are taken into account the system can become very complex.
Perhaps, it is not entirely unlikely that a similar kind of phase structure 
might also arise in $c>1$ matrix models.

In addition to being on the border between $c<1$ and $c>1$ theories, there is
another reason why the matrix chain is interesting. 
It models, in the large $N$ limit, one-dimensional string theory compactified
on a circle of radius $R=1/\epsilon.$ When defined on a
discretized worldsheet this theory automatically contains 
vortices---topologically nontrivial configurations where the string winds 
around the target space circle as we follow the boundary of an elementary
worldsheet plaquette.  
These vortices are suppressed thermodynamically for large $R$ (or, 
equivalently, small $\epsilon$) but become favored as $R$ is decreased.
As a result, they induce a phase transition in very much the same way 
the vortices of the two-dimensional $O(2)$ model drive the 
Berezinsky--Kosterlitz--Thouless phase transition 
\REFkosterlitzthouless. In fact, this 
Kosterlitz--Thouless-type  transition
is precisely the transition at $\epsilon=\epsilon_{\rm cr}$ which
separates the $c=1$ and $c=0$ phases of our theory.

To explore the dynamics of the Kosterlitz--Thouless phase transition in
string theory is a longstanding problem 
\REFgrossklebanov\REFboulatovkazakov\REFparisielse. In this paper we shall
present a computation which permits a systematic treatment of the transition
region. In particular, we find that the eigenvalue density at the 
Kosterlitz--Thouless critical point has an unusual logarithmic
singularity\foot{This should not be confused with the logarithmic
singularity in the free energy of $c=1$ models where $\rho(x)\propto
|x|$ without any logarithms.} $\rho(x)\propto |x|/\log(1/\lambda|x|).$
However, we also find that the inclusion of lattice fluctuations
gives rise to new Kosterlitz--Thouless-type transitions which would not
occur on a regular flat lattice. These are the transitions at $\epsilon=
\epsilon_{\rm cr}\sin(\pi p/2q).$ They arise when the fluctuating random 
lattice develops a curvature defect somewhere around the vortex core.
If such a defect introduces negative curvature, the energy of a vortex
decreases. As a consequence, the effective temperature of the 
Kosterlitz--Thouless transition or, equivalently, the critical value of
$\epsilon$ for such vortices becomes lower than $\epsilon_{\rm cr}$.

It is necessary to emphasize that these additional 
transitions do not modify the
double scaling limit of the matrix chain. Therefore, they do not correspond to 
any new phase transitions in continuum one-dimensional string theory.
However, they do exhibit universality properties and this makes them worthy
of consideration. Indeed, there are examples of phase transitions which appear
as lattice artifacts in one model but play an important role in the continuum
limit of another. For instance, the Gross--Witten phase transition in the 
one-plaquette model \REFgrosswitten, which represents 
a lattice effect, is in the same 
universality class with the Douglas--Kazakov  
phase transition \REFdouglaskazakov\ occuring
in continuum large $N$ Yang--Mills theory on a two-dimensional sphere.

Since the analysis presented below is somewhat lengthy we shall first
summarize the main steps of our computations and state the results.
This shall be done in the next section. In sections 3 and 4 we shall
present the details. In section 5 we check our results
by studying  the case of a special interaction
potential $V(M)$ for which the model can be solved exactly
in terms of elementary
functions. Finally, the vortex interpretation of defect-induced phase 
transitions shall be discussed in section 6.

\newsec{Large $N$ Expansion of Infinite Matrix Chain}

The free energy of the infinite one-dimensional random matrix chain 
represents, in the large $N$ expansion, the string perturbation
series for one-dimensional string theory with a discrete target 
space \REFparisioriginal\REFgrossklebanov.
To see this consider a cubic $V(M)=m^2 M^2/2 + g M^3/3$ and expand \chain\
in powers of $g$. Such expansion is a sum of all Feynman diagrams with
$\varphi^3$ vertices and propagators which can be inferred directly
from \chain,
\eqn\prop{
\big\langle M_{n,\alpha\beta} M_{n^\prime, \gamma\delta}\big\rangle=
N^{-1}\delta_{\beta\gamma}\delta_{\alpha\delta}D(n-n^\prime),}
where $\alpha,\beta,\gamma$ and $\delta$ are matrix indices
and the coordinate space propagator $D(n-n^\prime)$ is given by
\eqn\dprop{
D(n-n^\prime)=
\int \limits_{-\pi}^{\pi} {dp\over 2\pi} \thinspace{\rm e}^{ip(n-n^\prime)}
{\epsilon\over \epsilon^2 m^2 + 4 \sin^2 (p/2)}.}
According to these rules, 
the contribution of any individual Feynman graph $\Gamma$ with $V$ vertices
and $G$ handles
has the form 
\eqn\graph{
g^V N^{2-2G} \sum\limits_{n_1, \dots, n_V=-\infty}^{+\infty}
\prod\limits_{\langle ij\rangle} D(n_i-n_j).}
As usual, the product of propagators goes over
all links of the graph $\langle ij\rangle$ 
whereas the integers $n_1,\dots,n_V$ refer to
the vertex positions in the one-dimensional discrete coordinate space.
In the string theory language these integers parametrize the 
embedding of the graph $\Gamma$ into the discretised target space 
of the string.

The weight assigned to a given graph in this theory, the product of 
propagators
$D(n_i-n_j)$, is not exactly equal to the discretized Polyakov string weight.
Indeed, to obtain the Polyakov weight $\exp(-S_P[n_i])$ with
\eqn\polyak{
S_P[n_i]={1\over 2}\sum_{\langle ij\rangle} \epsilon^2 (n_i-n_j)^2}
one would have to choose the propagator $D_P(n)=\exp(-\epsilon^2 n^2/2)$.
Such a propagator could be generated only by a nonlocal, hard-to-deal-with
matrix model. Fortunately, it is possible to 
argue \REFcequalsone\REFgrossreview\ that the models
with these two weights are in the same universality class. The reason is 
that both $D$ and $D_P$ have the same infrared behavior in momentum space,
${\tilde D}(p)\propto 1- p^2/\mu^2.$ 
Since in one dimensional theory there are no 
ultraviolet divergences the replacement of $D_P$ by $D$ modifies only the
short-distance, nonuniversal properties of the model. 

Alternatively, the large $N$ limit of the infinite matrix chain \chain\
provides  a description of one-dimensional bosonic string theory 
compactified on a circle \REFgrossklebanov. 
However, in contrast with the discretized 
target space picture, this second interpretation is restricted 
to the leading order of large $N$ expansion. 
That is to say, our matrix model reproduces correctly only the first term
of the genus expansion for the compactified string.
To explain why let us perform a 
duality transformation on the Feynman amplitude \graph. 
As a first step, insert the momentum space representation \dprop\
for each $D$ in \graph\ and do the sums over $m_i$ using the identity
$$\sum_{n=-\infty}^{+\infty}{\rm e}^{ipn}=2\pi \sum_{l=-\infty}^{+\infty}
\delta(p+2 \pi l).$$
The result is 
\eqn\dualgraph{
g^V N^{2-2G}\sum\limits_{l_1, \dots, l_V=-\infty}^{+\infty}
\;\prod\limits_{\langle ij\rangle}\;
\int\limits_{-\pi}^{\pi} {dp_{ij}\over 2\pi}\thinspace
{\tilde D}(p_{ij})
\prod\limits_{k=1}^V \big[2\pi \delta (p^k_{\rm tot}- 2\pi l_k)\big]}
where $p_{ij}$ stands for the  momentum flowing along the link 
$\langle ij\rangle$, ${\tilde D(p)}$ is the momentum space propagator 
and $p^k_{\rm tot}=\sum_j p_{kj}$
equals the sum of all momenta entering the vertex number $k$.
Note that the vertices conserve momentum only modulo $2\pi$.

If $\Gamma$ is topologically spherical 
it is easy to interpret \dualgraph\ in terms of compactified 
string theory. 
To this effect we replace the link momenta $p_{ij}$ in \dualgraph\
by the string variables $X_a$, to be defined shortly, which will be 
associated with the vertices of the dual
graph $\Gamma^*$. Since we consider a compactified string 
the variable $X$ must live on a circle, so that $X$ and $X+2\pi R n$
for any $n\in\IZ$ are all identified.

\ifig\dualgraphs{A typical Feynman graph $\Gamma$ generated by a matrix 
model (solid lines) and its dual $\Gamma^*$ (dashed lines). 
The discrete target space coordinate $n$
is associated with the vertices of $\Gamma$ while the compactified matter 
field $X$ corresponds to the vertices of $\Gamma^*$.}
{\epsfysize 2.5in\epsfbox{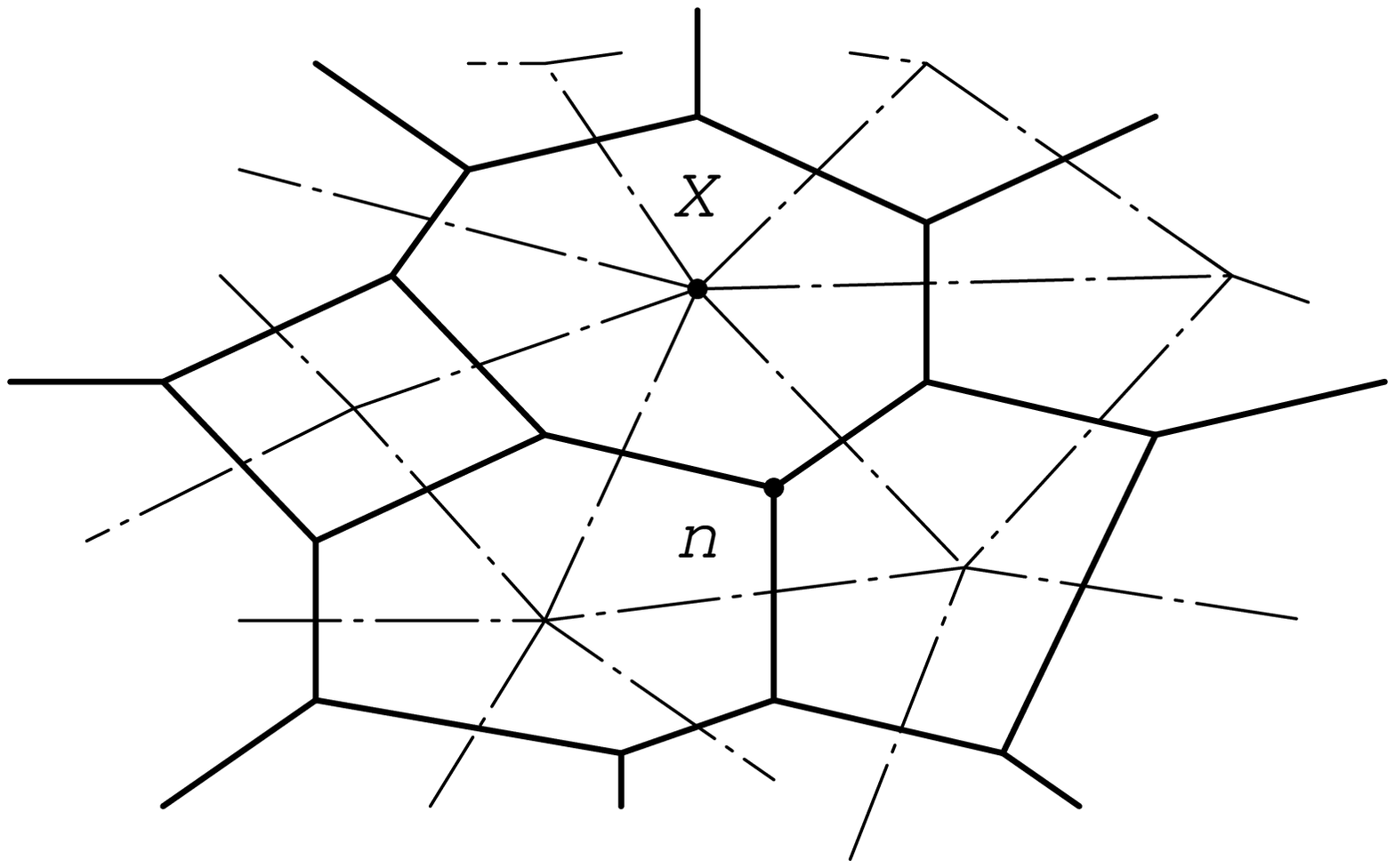}}

To construct $X_a$, for each link $\langle ij \rangle\in\Gamma$
we find a unique link $\langle ab \rangle$ of $\Gamma^*$ which
crosses $\langle ij \rangle$ and impose the condition
\eqn\duality{
X_a-X_b=R p_{ij}.}
If we fix the value of $X$ at any one vertex (say, set $X_1=0$) then equations
\duality\ define $X_a$ as functions of $p_{ij}$ at all other vertices.
In terms of these new variables \dualgraph\ takes the form
\eqn\compgr{
g^V N^2 \prod\limits_{a=2}^{V^*}\,
\int\limits_{-\pi R}^{\pi R} {dX_a\over 2\pi R}
\;\, \prod_{\langle bc\rangle} {\tilde D}\biggl({X_b-X_c\over R}\biggr)}
which looks very similar to \graph\ except that the target space is now a 
continuous circle of radius $R=1/\epsilon$.

The equivalence we just described does not hold on higher genus graphs
where in general \duality\ does not have a solution.  However, for our 
purposes this does not matter. Indeed, below we shall restrict
our attention to the large $N$ limit of the chain model where only the
spherical graphs survive and both interpretations apply.

To explore the large $N$ limit of the multimatrix chain we shall first 
integrate out the unitary degrees of freedom in \chain\ and then use 
the saddle point method. It will be convenient to rescale all matrices 
$M_n=\sqrt{\epsilon} {\cal M}_n$ so that 
$\epsilon$ disappears from the kinetic 
term in the partition function:
\eqn\chaintwo{
{\cal Z}=\int\prod\limits_{n=-\infty}^{\infty}\!\! d\calm_n
\; \exp\biggl\{N \, {\rm tr}\!\!\sum_{n=-\infty}^{\infty}
\Bigl[ \calm_n\calm_{n+1} - U(\calm_n)\Bigr]\biggr\}}
where 
the potential $U(\calm)$ is related to $V(M)$,
\eqn\potentials{
U(\calm)=\calm^2 +\epsilon V(\sqrt{\epsilon} \calm).}
As usual \REFbipz, we shall diagonalize each matrix, 
$\calm_n=U_n\Lambda_n U_n^{\dagger}$ with diagonal 
$\Lambda_n={\rm diag}(\lambda_{1,n}\dots \lambda_{N,n})$ and
express the Hermitian matrix measure $d\calm_n$ in terms of $U$'s and
lambdas,
$$d\calm_n=\Delta^2(\lambda_{n}) \thinspace d\lambda_{1,n} \dots 
d\lambda_{N,n} \thinspace dU_n$$
where $\Delta(\lambda_n)$ is the Van der Monde determinant of eigenvalues
$\lambda_{i,n}$ and $dU_n$ refers to the Haar measure on the unitary group
$SU(N)$. Furthermore, it is useful to introduce the matrices 
$V_n=U_{n+1}^{\dagger}U_n$.

Given this notation the partition function of the infinite matrix chain
can be written in the form 
\eqn\chaindiag{
{\cal Z}=\!\!\int\!\prod\limits_{n\in \IZ}^{}\!\!\Delta^2(\lambda_n)
 \, d\lambda_{1,n} \dots 
d\lambda_{N,n} dU_n \, \exp\biggl\{N \, {\rm tr}\!\!
\sum_{n=-\infty}^{\infty}
\Bigl[ V_n \Lambda_n V_n^{\dagger} \Lambda_{n+1} - U(\Lambda_n)\Bigr]\biggr\}.}
The matrices $V_n$ represent the angular degrees of freedom in our model.
Fortunately, since the one-dimensional lattice does not have closed loops,
all $V_n$ are mutually independent and we can easily integrate them out.
To do this one simply changes variables from $\{U_n\}$ to $\{V_n\}$ 
with the result
\eqn\chaindiagtwo{
{\cal Z}=\!\!\int\!\prod\limits_{n\in\IZ}^{}\Delta^2(\lambda_n)
\, d\lambda_{1,n} \dots 
d\lambda_{N,n} dV_n \, \exp\biggl\{N \, {\rm tr}\!\!\sum_{n=-\infty}^{\infty}
\Bigl[ V_n \Lambda_n V_n^{\dagger} \Lambda_{n+1} - U(\Lambda_n)\Bigr]\biggr\}.}
Note that no Jacobian arises when we pass from $U$ to $V$.

To integrate out $V$'s one must first compute the following
unitary group integral 
\eqn\izzub{
\int dV \; \exp\big[N \, {\rm tr} (V \Lambda_n V^{\dagger} \Lambda_{n+1})\big]=
\exp\big[ N^2 F(\Lambda_n,\Lambda_{n+1})\big].}
Then the resulting integral over the eigenvalues
\eqn\eigenint{
{\cal Z}=\!\!\int\! \prod\limits_{n\in\IZ}^{}\Delta^2(\lambda_n)
\, d\lambda_{1,n} \dots 
d\lambda_{N,n} dV_n \, \exp\biggl\{N \!\!\!\sum_{n=-\infty}^{\infty}
\Bigl[N F(\Lambda_n,\Lambda_{n+1})  - {\rm tr}\, U(\Lambda_n)\Bigr]\biggr\} }
can be evaluated, at $N\to\infty$, via the saddle point method.
In fact, due to the translational invariance of the matrix chain
the saddle point values of $\lambda_{i,n}$ are always independent of
the site number $n$. The saddle point equations which determine these
values are readily obtained by maximizing the integrand of \eigenint:
\eqn\saddlept{
2N {\partial F(\Lambda, \Lambda^{\prime})\over \partial \lambda_i}
\biggl|_{\Lambda^{\prime}=\Lambda} \!-{\partial U(\lambda_i)\over 
\partial\lambda_i} + {2\over N} \sum_{j\ne i} {1\over \lambda_i-\lambda_j}=0.}

The unitary integral \izzub\ has been computed in the classic paper 
of Itzykson and Zuber \REFitzyksonzuber. Unfortunately 
though, their result is not easy to  
use directly in the saddle point equation \saddlept. Indeed, 
the Itzykson--Zuber formula expresses $F$ through the determinant of a 
certain $N\times N$ matrix, a quantity which becomes quite complicated
in the large $N$ limit.

As $N$ goes to infinity it is convenient to characterize the eigenvalues
composing the diagonal matrices $\Lambda={\rm diag}(\lambda_1, \dots ,
\lambda_N)$ in terms of the so-called eigenvalue densities $\rho(\lambda)$.
By definition, $N\rho(\lambda)\, d\lambda $ is the number of those
eigenvalues among $\lambda_1, \dots ,\lambda_N$ which fall into an 
infinitesimal interval $[\lambda, \lambda+d\lambda]$. To define  the 
large $N$ limit of \izzub\ we enlarge diagonal matrices $\Lambda_n$ and
$\Lambda_{n+1}$  
in such a way that their respective densities of eigenvalues converge to the 
well defined
smooth limits $\rho_n(\lambda)$ and $\rho_{n+1}(\lambda)$. 
Note that as a
consequence of the above definition any density 
$\rho(\lambda)$ always obeys the
constraint 
$$\int\limits_{-\infty}^{+\infty} \rho(\lambda)\, d\lambda=1.$$

We shall now describe, without a derivation, how one computes $F$ for 
infinitely large $N$. It turns out that the large $N$ asymptotics 
of the Itzykson--Zuber integral \izzub\ is related to the classical
mechanics of a certain one-dimensional integrable system \REFhopf. This 
system is a 
droplet of one-dimensional compressible fluid which has the following peculiar 
equation of state
\eqn\eqnstate{
P=-{\pi^2\over 3}\rho^3}
where both the local pressure $P$ and the fluid density $\rho$ depend, 
in general, on  the
one-dimensional coordinate $x$.
The motion of such a droplet is described by Euler equations 
\eqn\euler{\eqalign{
&{\partial \rho\over \partial t}+{\partial \over \partial x}[\rho v]=0\cr
&\eqalign{{\partial v\over \partial t}+v {\partial v\over \partial x}
&=-{1\over \rho}{\partial P\over \partial x}\cr
&=\pi^2\rho{\partial \rho\over \partial x}\cr}\cr}}
where $v(x)$ is the local velocity at point $x$. 

To compute $F$ in equation \izzub\ one seeks the solution of these equations
which satisfies the boundary conditions
\eqn\bc{
\left\{\eqalign{
\rho(x, t=0)&=\rho_n(x)\cr
\rho(x, t=1)&=\rho_{n+1}(x)\cr}\right.}
where $\rho_n(x)$ and $\rho_{n+1}(x)$ are the eigenvalue densities 
for the matrices $\Lambda_n$ and $\Lambda_{n+1}$ respectively.
If the functions $\rho(x, t)$ and $v(x, t)$ yield such a solution
then
\eqn\F{
\eqalign{
&\lim\limits_{N\to\infty}F(\Lambda_n, \Lambda_{n+1})={1\over 2}
\int\limits_0^1 dt\int dx\, \rho(x, t)\biggl[v^2(x, t)+ {\pi^2\over 3}
\rho^2(x, t)\biggr]\cr
&+ {1\over 2}\int \rho_n(x) \, x^2\, dx 
-\int\!\int  dx_1 dx_2 \, \rho_n(x_1)\rho_n(x_2)\, \ln|x_1-x_2|\cr
&+ {1\over 2}\int \rho_{n+1}(y) \, y^2\, dy
-\int \!\int dy_1 dy_2 \, \rho_{n+1}(y_1)\rho_{n+1}(y_2)\, \ln|y_1-y_2|.\cr}}
It is worth mentioning that this expression for $F$ has a very special 
structure. Most importantly, the double integral 
\eqn\action{
S={1\over 2}
\int\limits_0^1 dt\int dx\, \rho(x, t)\biggl[v^2(x, t)+ {\pi^2\over 3}
\rho^2(x, t)\biggr]}
is actually the value of the classical action for the fluid droplet 
evolving according to equations \euler\ and \bc.

At this point we are ready to derive the large $N$ limit of the saddle 
point equation \saddlept. First of all, 
we need to know the derivative $\partial F(\Lambda_n, \Lambda_{n+1})/
\partial \lambda_i$ with respect to a given eigenvalue $\lambda_i$.
At large $N$ such a derivative can be easily expressed (simply by
manipulating the definitions) through the functional derivative
of $F[\rho_n, \rho_{n+1}]$ with respect to $\rho_n$,
\eqn\deriv{
N{\partial F(\Lambda_n, \Lambda_{n+1})\over \partial\lambda_i}=
\biggl({\partial\over \partial x}{\delta F[\rho_n, \rho_{n+1}]\over
\delta \rho_n(x)} \biggr)\biggr|_{x=\lambda_i}.}
Furthermore, using \action\ and equations of motion \euler\
it is possible to check that
\eqn\variationS{
{\partial\over \partial x}{\delta S[\rho_n, \rho_{n+1}]\over
\delta \rho_n(x)}=-v(x, t=0)}
and, therefore,
\eqn\variationF{
{\partial\over \partial x}{\delta F[\rho_n, \rho_{n+1}]\over
\delta \rho_n(x)}=-v(x, t=0)-x +2 -\kern-1.1em \int {\rho(y)\, dy\over x-y}.}
These relations are very natural. Indeed, the variation of the action $S$
on any classical solution with respect to the initial coordinate of that 
solution (in this case $\rho_n(x)$) always equals minus the corresponding 
canonical momentum.

We can use \deriv\ and \variationF\ to simplify the saddle point equation
\saddlept. Keeping in mind that
for $N\to\infty$
$${1\over N}\sum_{j\ne i}{1\over \lambda_i-\lambda_j}\rightarrow
-\kern-1.05em \int {\rho(y)\, dy\over \lambda_i-y} $$
one easily reduces the saddle point equation to
\eqn\spt{
v(x, t=0)={1\over 2} U^{\prime}(x)-x.}

The mathematical setup of the problem is therefore as follows.
Given the potential $U(x)$ one is looking for the solution of Euler equations
\euler\ which satisfies the boundary conditions
\eqn\bbc{
\left\{\eqalign{
&\rho(x, t=0)=\rho(x, t=1)\cr
&v(x, t=0)={1\over 2} U^{\prime}(x)-x.\cr}\right.}
Once this solution is found the saddle point density
$\rho(x, t=0)=\rho(x, t=1)\equiv\rho(x)$ can be used to compute ${\cal Z}$
or the free energy ${\cal F}=\log {\cal Z}$. Indeed, for the quartic
$U(x)=m^2 x^2/2 +gx^4/4$ it is easy to show that, as a consequence of
\eigenint\ and \saddlept\ 
\eqn\derF{
{\partial {\cal F}\over \partial g}= {N^2\over 4}\int \rho(x) \, x^4\, dx}
which immediately yields ${\cal F}(g)$ if $\rho(x)$ is known.

To say the same differently, imagine a fluid droplet of spatially
dependent density $\rho(x)$ (to be determined) being pushed with known
spatially dependent initial 
velocity $v(x)=U^{\prime}(x)/2-x$. Demand that after
one unit of time the density of that droplet evolves into the same $\rho(x)$.
This condition fixes $\rho(x)$, hopefully uniquely, for any given $v(x)$.

Most remarkably, Euler equations \euler\ are explicitly integrable.
This becomes clear if we combine $v$ and $\rho$ into
one complex-valued function $f=v+i\pi\rho$. One discovers that 
the two equations \euler\ are equivalent to the following complex
equation on $f$
\eqn\hopf{
{\partial f\over \partial t}+ f{\partial f\over \partial x}=0}
commonly known as the Hopf equation. 
The Cauchy problem for the Hopf equation is easily solvable. 
If $f_0(x)=f(x, t=0)$ is the initial value of the function $f$ then 
the value of $f(x, t)$ can be found from the implicit equation
\eqn\implicit{
f(x, t)=f_0\big[x-t f(x,t)\big].}
Now we can impose the boundary conditions \bbc\ in a more explicit form. 
By manipulating the implicit solution \implicit\ it is possible
to show  \REFhopf\ that \bbc\ gives rise to a functional equation on $\rho(x)$.
This functional equation is most easily formulated in terms of two auxiliary
functions 
\eqn\gpgm{
G_{\pm}(x)= {1\over 2}U^{\prime}(x) \pm i\pi\rho(x)}
and reads 
\eqn\gpmeq{
G_+\big[G_-(x)\big]=x.}
It provides a direct relation between the interaction potential
and the large $N$ eigenvalue density of the infinite matrix chain.

Unfortunately, little is known about solutions of such functional
equations \REFlittleknown. Certainly, it seems difficult to solve \gpmeq\
exactly for an arbitrary $U(x)$. This, however, is not always necessary.
Indeed, for most purposes one is interested only in the universal
properties  of $\rho$ --those independent of the particular $U(x)$
chosen.
For example, if a matrix model undergoes a phase transition the eigenvalue
density $\rho(x)$ usually develops a singularity at the transition
point. Such singularities are typically of the form $\rho(x)\propto
|x|^{\delta}$ (if the singular point is at $x=0$) characterized by a
universal exponent $\delta$. In fact, $\delta$ is closely related to the
string susceptibility $\gamma_{\rm str}=1-\delta$, a quantity which plays
an important role in string theory.

Therefore, one might try to evaluate such exponents without finding the
whole $\rho$. Luckily, equation \gpmeq\ is well suited for just that.
For concreteness and to make our problem well defined we shall concentrate
on even potentials bounded from below, such as 
\eqn\pott{
U(x)=-{\mu^2 x^2\over 2} +{g x^4\over 4}.}
Any other (say, cubic) potential can be treated the same way.
To set the notation let us choose $V(M)$ in the original definition
\chain\ as\foot{For a general quartic double well potential we can always 
rescale $M$ and $\epsilon$ 
in \chain\ to make the coefficient at $M^2$ equal $-2$. The physics of 
the model is not affected by such rescaling.}
\eqn\definvofm{V(M)=-2 M^2 +{{\tilde g}}M^4/4.}
This, according to \potentials\ means that 
\eqn\parameters{
\left\{\eqalign{
&\mu^2=2(2\epsilon^2-1)\cr
&g={\tilde g} \epsilon^3.\cr}\right.}
\ifig\densities{A typical double-well potential $V(x)$ 
and the corresponding eigenvalue density $\rho(x)$ in the $c=1$ 
matrix model at the critical coupling $g=g_{\rm cr}$
(thick solid  line), in the weak coupling phase $g<g_{\rm cr}$ 
(dotted line) and for strong coupling $g>g_{\rm cr}$ (dashed line).}
{\epsfxsize2.0in\epsfbox{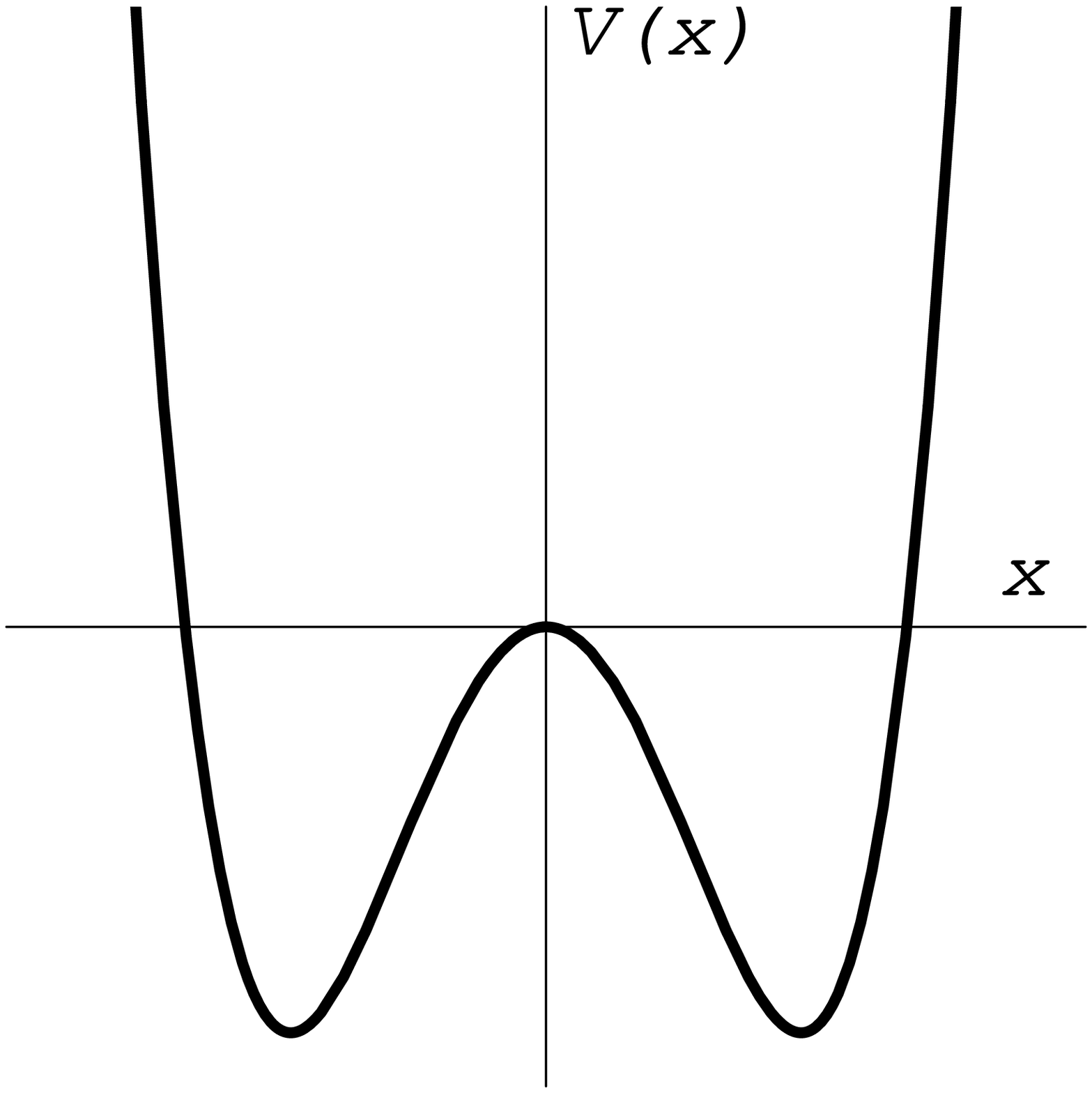}
\hskip0.5in\epsfxsize2.0in\epsfbox{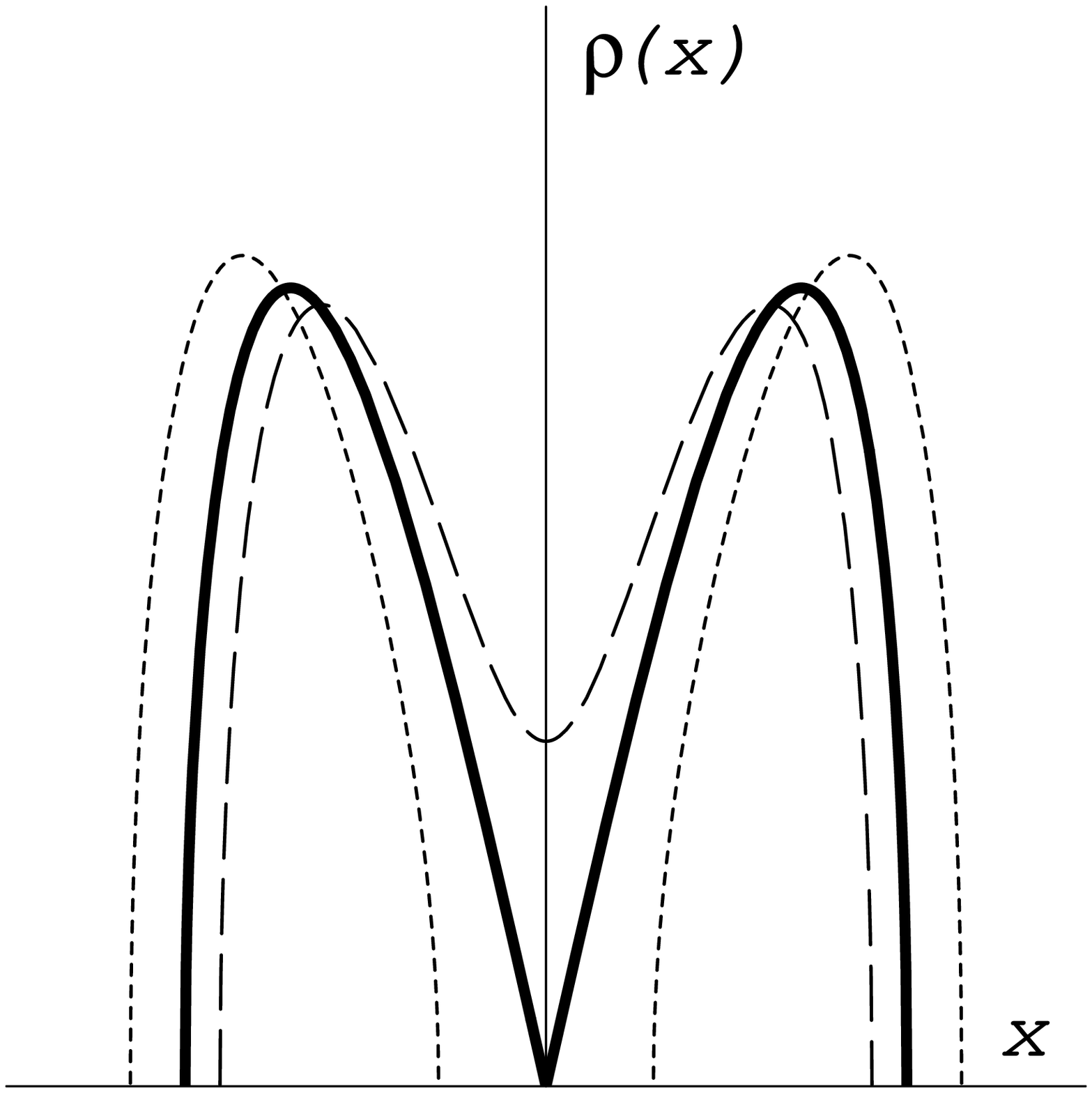}}
For any given fixed $\epsilon$ the multimatrix chain can be in one of the two
phases, depending on the value of $g$. For large $g$ (strong coupling)
the height of the hump in the potential $V(M)$ is relatively low 
so that the eigenvalue density $\rho(x)$ has only a small dip around $x=0$.
For very small $g$, however, the hump becomes high and the ``eigenvalue
droplet'' splits into two disconnected pieces. In between these two
phases at a certain $g=g_{\rm cr}(\epsilon)$
there is a phase transition where $\rho(x)$ just vanishes at
$x\to 0$,
\eqn\dell{\rho(x)\propto|x|^{\delta(\epsilon)}.}
The critical properties of the matrix chain encoded in $\delta(\epsilon)$
are not the same for various $\epsilon$. In fact, this model possesses
a number of multicritical points---special values of epsilon---where $\delta$ 
or other exponents can
abruptly change values. 
One of the problems we shall address below is to find and investigate
such multicritical points.

Before we proceed further it would be convenient to summarize our
findings:

(a) If $0<\epsilon<1$ then $\delta(\epsilon)=1$, an old result established 
originally by Gross and Klebanov \REFgrossklebanov. 
This value of $\delta$ is typical of $c=1$
theories. That is to say, the discrete structure of the target space
does not affect the string theory so long as the target space lattice
spacing is less than one.

(b) The Kosterlitz--Thouless 
point $\epsilon=1$ is special. It separates two regimes where the
infinite matrix chain has different values of $\delta$ and describes
different string theories. We find that for $\epsilon=1$ 
the critical eigenvalue density
(that is, the density at $g=g_{\rm cr}$) develops a logarithmic singularity
\eqn\logsing{
\rho(x)\propto {|x|\over \log\big[1/(\lambda|x|)\big]}.}
However, the universal properties of the matrix chain at the 
Kosterlitz--Thouless point are not exhausted by that. As we shall prove,
an infinite number of corrections to this logarithmic 
singularity also happen to be universal. More precisely, the
critical density $\rho(x)$ can be expanded in the form
\eqn\seriesuniv{
\rho(x)= |x|\, \, \sum_{q=0}^{\infty}\sum_{p=1}^q a_{pq}\,  l^p \, 
y^{-q-1} \, +\, 
{\cal O}(x^2)}
where $y\equiv \log[1/(\lambda|x|)]$ and $l\equiv\log y$.
The numbers $a_{pq}$ which arise as coefficients of this expansion
 are entirely independent of the interaction
potential $V(M)$. The potential enters only through 
the nonuniversal scale $\lambda$ and through the terms of order 
${\cal O}(x^2)$. 
We shall also see that the series in \seriesuniv\ can be summed. The result
is 
\eqn\sigg{
\rho(x)=|x|\, \sigma(x)}
where $\sigma(x)$ is determined by the transcendental equation
\eqn\transc{
{1\over \sigma}+\log\sigma =\log\biggl({1\over \lambda|x|}\biggr).}
Quite remarkably, nothing in this equation except for $\lambda$ depends on 
the matrix model potential. Therefore, 
equation \transc\ ought to have  a
continuum, string-theoretic interpretation. In the continuum language
the phase transition at $\epsilon=1$ is induced by topologically
nontrivial string configurations---vortices, which become strongly coupled 
for $\epsilon>1$. From this point of view \transc\ might contain 
information about vortex dynamics in the region where the ``vortex gas''
ceases to be dilute.


(c) Surprisingly, the linear matrix chain has other special points.
To observe them it is necessary to evaluate corrections to \dell. For a 
generic $\epsilon\in]0,1[$ these corrections are of the form 
\eqn\rhocorr{
\rho(x)=|x|\big\{a_1(\epsilon)+ a_2(\epsilon)x^2 + a_3(\epsilon)x^4 
+\dots\big\}}
where the coefficients $a_k(\epsilon)$ are fixed uniquely by Euler equations.
However, if $\epsilon=\sin(\pi p/2q)\equiv\epsilon_{pq}$ 
we discover that the coefficient $a_{q}(\epsilon)$
goes to infinity.  At these values of $\epsilon$ the corrections to $\rho(x)$
cannot be expanded in a power series. 
Indeed, we shall see that at $\epsilon=\epsilon_{pq}$
the correct expansion  for $\rho(x)$ contains 
a subleading  logarithmic singularity
\eqn\rhologsing{
\rho(x)=|x|\,\biggl\{a_1 +a_2 x^2 +\dots + a_{q-1}x^{2(q-2)}+a_q x^{2(q-1)}
\log\biggl({1\over \lambda |x|}\biggr)+\dots\biggr\}.}
These logarithmic singularities are universal. That is to say, they 
remain present if the quartic interaction is replaced by any other
generic polynomial potential $V(M)$. It is also possible to construct
a number of special, ``fine-tuned'' potentials 
for which some of the singularities get 
eliminated.

Mathematically, such logarithmic singularities are quite analogous to the 
Kosterlitz--Thouless singularity at $\epsilon=1$.
For example, if we shift $g$ away from $g_{\rm cr}(\epsilon)$ the shapes of
$\rho(x)$ at $\epsilon=1$ and at $\epsilon=\sin(\pi p/2q)$ are given by 
very similar expressions.
The qualitative pictures of what happens at these points seem to be 
related as well. Indeed, there is evidence that the singularities 
at $\epsilon=\epsilon_{pq}$,  like the Kosterlitz--Thouless phase 
transition at $\epsilon=1$, are due to effects of vortex proliferation.
Such proliferation is, of course, impossible for any $\epsilon<1$ 
in flat spacetime or on a regular lattice.  On a random lattice
the situation is different. There a pair consisting of a vortex and a 
negative curvature defect centered close to each other may become
a favored configuration even when $\epsilon<1$. We shall see that although
such pairs do not make a leading singular 
contribution to the eigenvalue density
or the free energy of the model, they are still sufficient to 
cause a mild singularity in $\rho(x)$.

(d) If $\epsilon>1$ and $V(M)$ is quartic the Euler equations are consistent
with $\delta(\epsilon)=2$. This is the same value of $\delta$ one observes
for $\epsilon\to\infty$ when the matrix chain becomes a collection 
of infinitely many decoupled one-matrix models. 
The transition between $\delta=1$ at $\epsilon<1$ to $\delta=2$ for 
$\epsilon>1$ has an interesting interpretation in terms of the 
hydrodynamic picture. It turns out that for $\epsilon>1$ the motion
of the liquid droplet prescribed by \euler\ and \bbc\ results in the 
formation of a shock. That is, at a certain moment in time, $t_{\rm sh}=
1/2\epsilon^2$ the spatial derivative of $\rho(x, t)$ at $x=0$ becomes
infinite. This phenomenon never happens for $\epsilon<1$. Moreover,
the droplet picture provides a natural order parameter for the 
Kosterlitz--Thouless transition. This order parameter
is defined by $\zeta(\epsilon)=\rho(x=0, t=1/2)$ where
$\rho(x, t)$ is the solution corresponding to $g=g_{\rm cr}(\epsilon)$.
The quantity $\zeta(\epsilon)$ has the property 
\eqn\orper{
\left\{\eqalign{
&\zeta(\epsilon)=0 \quad {\rm if}\quad \epsilon\le 1\cr
&\zeta(\epsilon)>0 \quad {\rm if}\quad \epsilon >1 \cr}\right.}
and will play an important role in our analysis of the $\epsilon >1$ phase.

(e) Finally, we shall study the character expansion of the 
infinite matrix chain. The idea is to expand the integrand of the 
partition function 
\chaindiagtwo\ in a certain sum over $U(N)$ characters \REFwynter. 
Then within each term
of this sum the unitary degrees of freedom $V_n$ can be integrated out 
with ease. At large $N$ the resulting sum over representations of $U(N)$ 
is dominated by a single ``saddle point representation.''
 The highest weights of this representation $(n_1, \dots, n_N)$ are 
all integers of order $N$. It is conventional to characterize such
 highest weights by the 
``Young tableau density'' $\rho_l(h)$ defined as the density of points 
$h_i=(n_i-i)/N+1/2$ in a small interval around $h$. 
Note that since the highest 
weights are ordered the Young tableau density can never be 
greater than one. 
We shall prove that the Young tableau density $\rho_l(h)$
is directly related to the density $\rho_{1/2}(x)=\rho(x, t=1/2)$
which arises in the hydrodynamic picture:
\eqn\dua{
\pi \rho_{1/2}\biggl[\sqrt{h}\,\cos\biggl({\pi \rho_l(h)\over 2}\biggr)\biggr]
=2\sqrt{h}\, \sin\biggl({\pi \rho_l(h)\over 2}\biggr).}
This relation shows that there is a lot in common between the 
Kosterlitz--Thouless phase transition and the Douglas--Kazakov-type 
transitions which are observed in two-dimensional QCD or in dually
weighted graph models \REFdouglaskazakov\REFwynter. In the 
Douglas--Kazakov transitions
the Young tableau density is always less than one for weak coupling but 
develops a plateau with
$\rho_l(h)=1$ for any $h\in[-h_*, h_*]$ at strong coupling.
Exactly the same happens in the matrix chain. Indeed, since for
$h=h_*$ the density $\rho_l(h_*)=1$, equation \dua\ predicts
\eqn\rell{
\zeta=\rho_{1/2}(0)={2\over \pi}\sqrt{h_*}.}
Equation \orper\ now  implies that the plateau in $\rho_l$
is present only for $\epsilon>1$, in complete analogy with the
Douglas--Kazakov transition.
The physical pictures of the two transitions are also analogous. 
Indeed, both the Kosterlitz--Thouless and the Douglas--Kazakov transitions
are induced by the topologically nontrivial configurations (vortices and 
instantons, respectively) \REFdouglaskazakov\REFgrossinstantons\
which dominate in the strong coupling phases of 
two-dimensional $O(2)$ model and two-dimensional QCD.

A more detailed analysis of the infinite matrix chain at $\epsilon>1$
will be given separately in another report. Below we shall discuss the 
Kosterlitz--Thouless phase transition and the subleading logarithmic 
singularities which arise in the $\epsilon\le 1$ phase of the model.

\newsec{Critical Behavior of Eigenvalue Densities for $\epsilon<1$}

In this section we shall explore the critical properties of the eigenvalue 
density for lattice spacings smaller than one.

Let us first concentrate on the case when the coupling constant $g$ is exactly
equal to the critical value $g_{\rm cr}(\epsilon)$. In the language of 
continuum theory this would correspond to setting the renormalized 
cosmological constant to zero.
Furthermore, since the singularity of $\rho(x)$ occurs at $x=0$ we can restrict
our attention to small values of $x$. 

For quartic $U(x)$ given by \pott\ and 
\parameters\ the boundary conditions \bbc\
read 
\eqn\boundcond{\eqalign{
v(x, t=0)\equiv v_0(x)&={1\over 2} U^{\prime}(x)-x \cr
&=-2\epsilon^2 x +{g x^3\over 2}.\cr}}
Naively, when $x$ is small we can neglect the cubic term in $v$ and 
consider the simplified boundary problem where the initial velocity is given 
by
\eqn\vsimpl{v_0(x)=-2\epsilon^2 x.}
The solution of Euler equations subject to this boundary condition is easy to
construct. We simply observe that the ansatz
\eqn\ansatz{
\left\{\eqalign{
\pi\rho(x, t)&=\alpha(t) |x|\cr
v(x, t)&=\beta(t) \, x\cr}\right.}
is consistent with Euler equations \euler\ provided 
\eqn\time{
\left\{\eqalign{
&{\dot \alpha}+ 2 \alpha\beta=0\cr
&{\dot \beta}+ \beta^2 = \alpha^2\cr}\right.}
where the dot denotes the time derivative.

These equations can be easily solved. One introduces the complex valued 
function $f_1(t)=\beta(t)+i\alpha(t)$ which, as a consequence of \time\
obeys 
\eqn\fonedot{
{\dot f_1}+f_1^2 =0,}
so that
\eqn\fonedotsol{
f_1(t)={1\over t- c_0}}
with a certain complex valued constant $c_0$.
This constant should be determined from the boundary conditions. The first 
condition in \bbc\ amounts to ${\rm Im}f_1(0)={\rm Im}f_1(1)$ which entails
$$c_0={1\over 2}+i\tau_0 $$
with real $\tau_0$. Then from the condition on the initial velocity,
${\rm Re}f_1(0)=-2\epsilon^2$ one determines the constant $\tau_0$
\eqn\tauzero{
\tau_0={1\over 2\epsilon}\sqrt{1-\epsilon^2}}
and, finally, the eigenvalue density at small values of $x$
\eqn\rhoansw{\eqalign{
\pi\rho(x, t=0)&=\alpha(0) |x|\cr
&=2\epsilon\sqrt{1-\epsilon^2} \, |x|.\cr}}
The critical exponent for such $\rho$ is obviously
$\delta(\epsilon)=1$. Furthermore, 
we see that the expression for $\rho(x)$ given by \rhoansw\ degenerates
when $\epsilon=1$. This is the first sign of the Kosterlitz--Thouless phase 
transition which occurs at that point.

The simple solution we just constructed exhibits several important general
features. First, it is symmetric with respect to time reflection
around $t=1/2$:
\eqn\timerefl{
\left\{\eqalign{
&\rho(x, 1-t)=\rho(x, t)\cr
&v(x, 1-t)=-v(x, t).\cr}\right.}
This property is in fact true for any solution of Euler equations which obeys
boundary conditions \bbc.
The reason is that both 
Euler equations and the boundary conditions 
are invariant with respect to the reversal of time. 
That is to say, if $\rho(x, t)$ and $v(x, t)$ yield a solution, so do
${\tilde \rho}(x, t)=\rho(x, 1-t)$ and ${\tilde v}(x, t)=-v(x, 1-t)$.
Furthermore, due to the translational invariance of the matrix chain
the final velocity at $t=1$ is opposite to the initial velocity $v_0(x)$,
\eqn\oppos{v(x, t=1)= -v(x, t=0).} This, together with $\rho(x, t=0)=
\rho(x, t=1)$, implies that the solutions
$\{{\tilde \rho},{\tilde v}\}$ and $\{\rho, v\}$ coincide at $t=0$.
Consequently, these two solutions coincide also at all
later times, and \timerefl\ follows.

The relation between the initial and final
velocities, which we needed for the proof, is almost
obvious.
Indeed, 
the action $S[\rho_n, \rho_{n+1}]$ is a symmetric functional
of the densities $\rho_n$ and $\rho_{n+1}$. Therefore if $\rho_n=\rho_{n+1}$
the functional variation of $S$ with respect to $\rho_n$ equals  
the variation of $S$ with respect to $\rho_{n+1}$. On the other hand, the end
velocity is related to the variation of the action by 
\eqn\variationSS{
{\partial\over \partial x}{\delta S[\rho_n, \rho_{n+1}]\over
\delta \rho_{n+1}(x)}=+v(x, t=1)}
with a plus sign. On comparison with equation \variationS\ we 
immediately deduce
\oppos.

An important consequence of the time reflection symmetry is 
\eqn\vhalf{v(x, t=1/2)=0.}
In other words, after one half unit of time the fluid completely stops moving.
This leads to the following qualitative picture of the droplet evolution.
At the initial moment of time $t=0$ the fluid in the vicinity of $x=0$ 
is moving inwards. The pressure $P=-\pi^2 \rho^3/3$ acts to slow down and
stop this motion. If the initial $\rho(x)$ has been chosen properly then
at $t=1/2$ the velocity of the fluid vanishes simultaneously everywhere.
At larger $t$ the forces of pressure make the droplet move again, repeating
backwards in time the evolution from $t=0$ to $t=1/2$. 

Let us now see how the functional equation \gpmeq\ reproduces the result of
\rhoansw. This equation involves the density of eigenvalues taken at a
complex point, $\rho[G_-(x)]$.
Such an object is certainly well defined if $\rho(x)$ is analytic. However,
the density profile of interest to us, $\rho(x)\propto|x|$ cannot be continued 
analytically into the complex plane. 

The resolution of the arising difficulty is quite simple. One should throw 
away the absolute value sign and consider instead $\rho(x)\propto x$. Such 
prescription does yield the correct answer. Indeed, if $\pi\rho(x)=\alpha x$
the functions $G_+$ and $G_-$ are given by
\eqn\gggg{
G_{\pm}(x)=(1-2\epsilon^2 \pm i \alpha)x.}
Then equation \gpmeq\
reduces to
$$(1-2\epsilon^2)^2 +\alpha^2=1$$
which predicts the correct value of $\alpha$
$$\alpha=2 \epsilon\sqrt{1-\epsilon^2}.$$
This prescription works because no fluid ever flows through $x=0$.
In fact, the regions $x>0$ and $x<0$ do not really interact with each other. 
Furthermore,
nothing in Euler equations requires the density to be manifestly positive.
For instance, we can flip the sign of $\rho(x,t)$ for $x<0$. The density
so obtained, $\rho(x,t)\propto x$ 
will be analytic in $x$ and will still satisfy Euler equations.
On the other hand, the $x>0$ part of the eigenvalue density remains unchanged 
allowing us to read off the correct answer.

So far the quartic interaction $g \, {\rm tr}M^4/4$ has not played any role 
in our analysis. An identical effect occurs in the standard formalism 
of $c=1$ string theory based on matrix 
quantum mechanics \REFcequalsone\REFgrossklebanov.
There, too, one expands the potential
around the local maximum and only the quadratic terms matter.
In the hydrodynamic picture  the interaction terms modify the initial 
velocity of the droplet by an amount negligible for small $x$. Therefore,
for $x\to 0$ the effects of interaction can be taken into account
via perturbation theory. The validity of such perturbative expansion 
is ensured by the smallness of $x$ even though the coupling $g$ is of 
order one. In a sense, we are lucky that this  perturbation theory 
is valid precisely in the critical domain.

When the lattice spacing gets bigger than one such a simple picture 
ceases to be valid. For $\epsilon>1$ imparting the initial velocity
$v=-2\epsilon^2 x$ to a droplet configuration $\rho(x)\propto |x|$ will always 
make the droplet collapse. 
In this case the cubic correction to the initial $v$ given by $gx^3/2$
provides a small, but essential, amount of outward directed velocity that
is needed to prevent such  collapse.


Quite surprisingly, the interaction effects are not completely trivial
even if $\epsilon$ is less than one. 
This can be seen already in the first order
of perturbation theory. Indeed, let us evaluate the first correction
to the linear ansatz \ansatz. To this end we expand $\rho$ and $v$ in powers
of $x$
\eqn\exppowers{
\left\{\eqalign{
\pi\rho(x, t)&= \alpha_1(t)|x|+\alpha_2(t)|x|^3+\dots\cr
v(x, t)&=\beta_1(t)\, x +\beta_2(t)\, x^3+\dots\cr}\right.}
Technically, it is more convenient to work directly with the Hopf
function $f=v+i\pi\rho$. For $x>0$ this function has a power
series expansion in $x$,
\eqn\hopfexpP{
f(x, t)=f_1(t)\, x + f_2(t)\, x^3+\dots.}
Substituting this into the Hopf equation \hopf\ and collecting the terms 
of order $x^3$ we obtain an ordinary differential equation on $f_2(t)$
\eqn\ftwo{
{\dot f_2} +4 f_2 f_1=0.}
Now we can use the explicit solution for $f_1$ given by \fonedotsol\ to find
\eqn\ftwosol{
f_2(t)= {i c_2\over (\tau-i\tau_0)^4}}
where $\tau\equiv t-1/2$ and $c_2$ is a constant of integration. At $\tau=0$
we must have $v=0$ which means that $c_2$ is real. Further, the
boundary condition \boundcond\ yields ${\rm Re}f_2(t=0)=g/2$
fixing $c_2$ to be
\eqn\ctwo{
c_2=-{g\over 32 \epsilon^4 \sin(4\arcsin\epsilon)}={g\over 128 \epsilon^5 
(2\epsilon^2-1)\sqrt{1-\epsilon^2}}.}
As a result, we find the first correction to the eigenvalue density
\eqn\rhocorrfirst{
\pi\rho(x)= 2\epsilon\sqrt{1-\epsilon^2}\, |x| -{g |x|^3\over 2 \tan
(4\arcsin\epsilon)}+\dots.}
If $\epsilon=1/\sqrt{2}$ this correction becomes infinite. The technical
reason for that is very clear. When $\epsilon=1/\sqrt{2}$ the value of 
$f_2(t)$ at $t=0$ is purely imaginary, $f_2(0)=-4ic_2$. Therefore,
even though we do perturb the eigenvalue density by $x^3$ terms, no change 
in the initial velocity arises. In this way, the $x^3$ mode at 
$\epsilon=1/\sqrt{2}$ could be viewed as a ``resonance'' of the droplet.
Certainly, a similar picture of resonances should also appear
in the equivalent approach to $c=1$ strings based on
matrix quantum mechanics.

Whenever such resonances occur the perturbative expansion needs to be 
modified. It turns out that an adequate modification is given by the 
following formula
\eqn\modif{
f(x, t)=f_1(t)\, x + f_2(t) \, x^3\log x +{\tilde f}_2(t)\, x^3 +\dots.}
For this to be consistent with the Hopf equation $f_2(t)$ 
must satisfy the old differential equation \ftwo\ while ${\tilde f_2}(t)$
should obey
\eqn\ftwotilde{
{\dot{\tilde f_2}}+4 f_1 {\tilde f_2} + f_1 f_2 =0.}
Consequently, the function $f_2(t)$ is still
given by formula \ftwosol\ with a real
$c_2$ and $\tau_0=\sqrt{1-\epsilon^2}/2\epsilon=1/2$. However, 
the constant $c_2$
is not fixed by the boundary conditions on $f_2(t)$ anymore. Instead,
it will be determined from the boundary conditions on ${\tilde f}_2$.

Since the initial velocity $v_0(x)$ contains no logarithms we must
demand that ${\rm Re}f_2(t=0)=0$. According to \ftwosol\ this is possible
only if $\epsilon=1/\sqrt{2}$. In other words, it is not always possible
to add $x^3\log x$ to the eigenvalue density---for that the lattice spacing
must take a special value. And, in agreement with the picture of resonances,
it is precisely at this special value when the original power series
\rhocorrfirst\ fails.

Given an explicit solution for $f_2$ (with $c_2$ undetermined) we can find 
${\tilde f}_2$ from \ftwotilde.
Imposing the usual conditions
${\rm Re}{\tilde f}_2(t=1/2)=0$ and ${\rm Re}{\tilde f}_2(t=0)=g/2$
fixes $c_2$ in \ftwosol\ to be $c_2=g/2\pi$ and
leads to the following expression for ${\tilde f}_2(t)$
\eqn\ftildesol{
{\tilde f}_2(t)={i\over\Bigl({1\over 2}+i \tau\Bigr)^4}\biggl[
-{g\over 2\pi}\log\biggl({1\over 2}+i \tau\biggr) + {\tilde c_2}\biggr]}
with a yet undetermined constant ${\tilde c_2}$.
The eigenvalue density which corresponds to such $f$
has a logarithmic singularity at $x=0$
\eqn\rhosollog{
\pi\rho(x)=|x|+ {2g\over \pi}|x|^3 \log\biggl({1\over \lambda |x|}\biggr)+
{\cal O}(x^4).}
The scale $\lambda$ is related to ${\tilde c_2}$ by ${\tilde c_2}=
(g/4\pi)\log(\lambda^2/2)$ and cannot be determined from perturbation theory.
To find such a scale, as well as the specific numerical 
value of the critical coupling
$g=g_{\rm cr}(\epsilon)$ one would have to construct 
the eigenvalue density 
for all, not only small, $x$
and impose 
certain conditions on the global analytic structure and normalization of 
$\rho(x)$. These conditions shall be discussed in more detail at the end 
of the next section. 
However, notwithstanding any ambiguity in the value of ${\tilde c_2}$
the structure of the singularity in $\rho(x)$
is still
fixed uniquely by the Hopf equation.

Similar logarithmic singularities arise also in higher orders. 
In general, for $\epsilon=\epsilon_{pq}\equiv\sin(\pi p/2q)$
the eigenvalue density has a logarithmic correction of order 
$|x|^{2q-1}\log[1/(\lambda|x|)]$. This can be proved in a rather
elementary fashion with the aid of the functional equation \gpmeq.
Indeed, if the eigenvalue density does not have logarithms the
functions $G_{\pm}(x)$ can be expanded in a power series
\eqn\gpser{
G_{\pm}(x)=\sum_{k=1}^{\infty}b^{\pm}_k x^{2k-1}.}
By construction, $G_+$ and $G_-$ are complex conjugate, so that
$b_k^+=(b_k^-)^*$. Furthermore, since ${\rm Re}\, G_{\pm}(x)=U^{\prime}(x)/2$
is a cubic polynomial all $b_k^{\pm}$ with $k\ge 3$ are purely imaginary
and thus satisfy $b_k^-=-b_k^+$.

Let us try to solve the functional equation \gpmeq\ order by order in $x$.
In the first order, $b_1^+ b_1^-=1$ which means $b_1^{\pm}=\exp(\pm i\varphi)$
with a certain real $\varphi$. Actually, comparing \gpser\ to 
\gggg\ we see that the angle $\varphi$ is 
related to the lattice spacing $\epsilon$ by $\cos\varphi=1-2\epsilon^2$
or, equivalently, $\epsilon=\sin(\varphi/2)$.

If we know all $b_k^{\pm}$ with $k\le q-1$ and wish to find $b_q^{\pm}$
we expand $G_+[G_-(x)]$ in powers of $x$ and collect the terms of order
$x^{2q-1}$. As a consequence of \gpmeq\ the sum of such terms must vanish.
This yields
\eqn\vanishsum{
b_q^+ (b_1^-)^{2q-1} + b_1^+ b_q^- + \left\{{{\rm terms\  which\  depend} 
\atop
{\rm only\ on}\ b_1^{\pm}, \dots, b_{q-1}^{\pm}}\right\}=0.}
For certain values of $\epsilon$  equation \vanishsum\ 
degenerates. Indeed, since
$b_q^-=-b_q^+$ and $b_1^{\pm}=\exp(\pm i\varphi)$ we find that 
$$b_q^+ (b_1^-)^{2q-1} + b_1^+ b_q^-=-2i b_q^+ \sin(q\varphi)\, 
{\rm e}^{-i(q-1)\varphi}.$$
If $\varphi=\pi p/q$ the variable $b_q^+$ disappears from the right hand side 
of \vanishsum. Then, unless the rest of \vanishsum\ vanishes at the same time,
we encounter a contradiction. As a consequence, for these values of $\varphi$
or, equivalently, for $\epsilon=\sin(\pi p/2q)$ 
the eigenvalue density develops an additional singularity. We have checked 
explicitly that this  is indeed true in the first 
seven orders of perturbation theory.

The same type of reasoning demonstrates that the position, the order 
and the logarithmic 
nature of arising singularities are preserved if the quartic interaction
$V(M)$ is replaced by a more complicated potential.
There may be, however, exceptional situations. They occur when the terms
enclosed by the curly brackets in equation \vanishsum\ vanish simultaneously
with $\sin(q\varphi)$. For that to happen the coefficients of the matrix 
potential $V(M)$ must be adjusted in a special way. 
The eigenvalue density which corresponds to such fine-tuned, ``multicritical''
potentials will exhibit only some, or none, of the above singularities.
For instance, in section 5 we shall construct an example of a nonpolynomial
potential which gives rise to only one singularity at $\epsilon=1/\sqrt{2}$.
The model with such a potential can be solved exactly. In fact, it is
a generalization of the Penner-type model where an exact solution
can be gotten independently using the method of loop equations \REFpenner.

The above considerations apply when the coupling constant $g$ strictly
equals $g_{\rm cr}(\epsilon)$. Only then $\rho(x)\propto|x|$ around $x=0$.
Whenever $g$ is shifted away from $g_{\rm cr}(\epsilon)$ (or, in the language 
of continuum theory, when the cosmological constant is given a nonzero value)
the eigenvalue 
density gets perturbed.
Such perturbed shape of $\rho(x)$ would be useful to know for several reasons.
First, if one wishes to explore the critical behavior of the free energy,
say using equation \derF, one needs to know $\rho(x)$ in a certain interval
of couplings around $g_{\rm cr}(\epsilon)$. Further, we shall see that
such analysis provides an independent derivation of the 
logarithmic singularities
which were simply guessed in equation \modif.

For a 
small nonzero $\delta g=g-g_{\rm cr}(\epsilon)$ the $|x|$-type singularity
in $\rho(x)$ disappears. Instead the eigenvalue density develops a pair of 
close square root branch points. These branch points merge when $\delta g$ 
vanishes producing $\rho(x)\propto |x|$. 
In fact, the analytic structure of exactly the same kind occurs 
in the $c=1$ matrix model with a continuous target space 
where the eigenvalue density is
given by the WKB formula \REFcequalsone\REFgrossinducedQCD
\eqn\wkb{
\pi\rho(x)=\sqrt{2\bigl[E-{\tilde U}(x)\bigr]}}
with ${\tilde U}(x)=-{\tilde \kappa}^2 x^2/2+gx^4/4$.

Remarkably, the existence of two square root branch points is fully 
consistent with Euler equations. To see this let us consider the following
``hyperbolic'' ansatz for $\rho$ and $v$
\eqn\hyper{
\left\{\eqalign{
\pi\rho(x, t)&=\sqrt{b(t)+\alpha^2(t) \, x^2}\cr
v(x, t)&=\beta(t)\, x\cr}\right.}
This ansatz is consistent with Euler equations provided that
$\alpha(t)$ and $\beta(t)$ obey the old equations \time\ while $b(t)$
satisfies
\eqn\boft{
{\dot b}+2 \beta b=0.}
Actually, the linear ansatz \ansatz\ that we considered previously 
is simply a $b=0$ case of the hyperbolic solution.

The eigenvalue density which correspond to \hyper\ 
\eqn\rhohype{
\pi\rho(x)=\sqrt{b(0)+ 4\epsilon^2(1-\epsilon^2)x^2}}
has two square root branch points. Moreover, such $\rho(x)$ can be viewed 
as a generalization of the WKB formula \wkb\ to finite nonzero lattice 
spacings. Indeed, for small $x$ the WKB solution has the form 
$$\pi\rho(x)=\sqrt{2 E+ {\tilde \kappa}^2 x^2}$$
identical to \rhohype.
The parameter $b\equiv b(0)$ in the 
hyperbolic solution and the Fermi energy $E$
in \wkb\ play extremely similar roles. Both $b$ and $E$ are in fact functions 
of $\delta g$ which are determined by the normalization condition that
the total integral of the full, exact $\rho(x)$ equals one.
Obviously, for small coupling deviations $\delta g$ the parameters 
$b$ and $E$ are also small.

In the critical region of interest to us both $b$ and the eigenvalue magnitude
$x$ are small but can be 
of the same order with respect to each other. That is, if the characteristic
width of the double well potential $V(M)$ equals $a$ then $x\sim \sqrt{b}
 \ll a$. It is exactly this region of eigenvalues---a small interval
of size $\sim \sqrt{b}$ around the top of the potential---that is important,
for example, in the matrix quantum mechanics representation of $c=1$ strings.
For such $x$ and $b$ equation \rhohype\ yields a good approximation
to $\rho(x)$ which can be systematically improved upon.

The corrections to \rhohype\ can be easily computed using the functional
equation $G_+[G_-(x)]=x$. We shall look for the corrected eigenvalue
density in the form
\eqn\rhocorrbnotzero{
\pi\rho(x)=\sqrt{b+ x^2 \sin^2\varphi +g r(x)}}
where $\varphi$ is defined by $\epsilon=\sin(\varphi/2)$ and $r(x)$ 
represents the first correction. For $x^2\sim b$ the quantity $r(x)$ 
should be small of order $b^2$ or, equivalently, $x^4$.
Actually, for the special case of $b=0$ the estimate $r(x)\sim x^4$ follows 
already from \rhocorrfirst. The functions $G_+$ and $G_-$ are then
\eqn\gplusminus{
G_{\pm}(x)=x \cos\varphi +{g x^3\over 2}+i\sqrt{ b+ x^2 \sin^2 \varphi + 
g r(x)}.}
The approximation we are constructing is formally the same as the 
power series expansion in $g$. Indeed, in such an expansion each extra 
power of $g$ comes along with either $b$ or $x^2$. For example, $g r(x)\sim
g b^2$ is much smaller than $b+ x^2 \sin^2 \varphi$ which is of order $b$. 
The same applies to $gx^3/2\sim gb x$ which is much smaller than 
$x\cos\varphi$.
In short, the true parameter of our expansion is $gb\sim gx^2\ll 1$ so 
that we can formally expand in $g$ even though $g$ by itself is not small.

Doing this in \gplusminus\ to first order in $g$ yields 
$$G_{\pm}(x)=G_{\pm}^{(0)}(x) +g H_{\pm}(x)$$
with 
\eqn\gpmzeorord{
G_{\pm}^{(0)}(x)=x\cos\varphi \pm i \sqrt{b+ x^2 \sin^2 \varphi}}
and 
\eqn\hpm{
H_{\pm}(x)={x^3\over 2} \pm i {r(x)\over 2\sqrt{b+ x^2 \sin^2 \varphi}}.}
Then the functional equation \gpmeq\ expanded to ${\cal O}(g^2)$ reduces to
\eqn\gper{
G_+^{\prime(0)}\big[G_-^{(0)}(x)\big] H_-(x) + H_+\big[G_-^{(0)}(x)\big]=0.}
We now substitute for $G_{\pm}^{(0)}$ and $H_{\pm}$ the explicit expressions 
of \gpmzeorord\ and \hpm.
This produces the following linear functional equation on $r(x)$
\eqn\linfun{
r(y)-r(x)=xy^3-x^3 y +(x^4-y^4)\cos\varphi}
where the letter $y$ stands for the combination
$$y\equiv x \cos\varphi-i\sqrt{b+ x^2 \sin^2 \varphi}.$$ 
When $x^2\sim y^2\sim b$ both sides of this equation are of order $b^2$. 
It is therefore natural to look for a solution of \linfun\ in terms of a 
polynomial ``homogeneous'' in $x^2$ and $b$
\eqn\hompol{
r(x)= R_1 x^2 b+ R_2 x^4.}
Upon a certain amount of computation one finds that this form of $r$ is 
indeed consistent with the functional equation \linfun\ provided $R_1$ and
$R_2$ equal
\eqn\Rbig{
\eqalign{R_1&={1\over 2 \cos\varphi}={1\over 2(1-2\epsilon^2)};\cr
R_2&= {1\over 2 \cos\varphi}-\cos\varphi={1\over 2(1-2\epsilon^2)}-1+2
\epsilon^2.\cr}}
Expressions \hompol\ and \Rbig\ generalize the 
perturbative expansion \rhocorrfirst\ to the case of a nonzero cosmological 
constant. Not surprisingly, these expressions also develop 
poles at $\epsilon=1/\sqrt{2}$, where the polynomial form for
$r(x)$ ceases to be a solution of \linfun.
However, even for $\epsilon=1/\sqrt{2}$ the equation on $r(x)$ can still be
solved. 
Remarkably, the solution will produce a logarithmic contribution in $r$
automatically without adding it by hand as we did before in \modif.

When $\epsilon=1/\sqrt{2}$ our linear functional equation becomes
\eqn\linfunpole{
\left\{\eqalign{
& r(y)-r(x)= y^3 x -x^3 y\cr
& y\equiv -i\sqrt{b+x^2}\cr}\right.}
We are interested in the solution of this equation which has two square 
root branch points at $x=\pm i \sqrt{b}$ and, on the real axis, is an even 
function of $x$ regular at $x=0$.

Such a solution is defined on a complex plane with two cuts as shown in the
figure and can be found using the method of dispersion relations\foot{A
shorter way to solve this equation, suggested by J. Goldstone, is to
use the substitution $x=\sqrt{b} \,\sinh\psi.$ Nonetheless, the solution based 
on dispersion relations will be necessary to illustrate several important 
points later, in section 4.}. 
Indeed, the imaginary part of $r(x)$ on the edges of the cuts follows directly
from \linfunpole\ and equals
\eqn\impart{
{\rm Im}\, r(-iu+0)= u (2u^2-b)\sqrt{u^2-b}.}
To see this, consider equation \linfunpole\ with a positive real $x$. Then 
$y=-iu=-i\sqrt{b+x^2}$ is positioned on the right edge of the lower cut.
Furthermore, ${\rm Im}\, r(x)=0$. As a consequence, ${\rm Im}\, r(y)=
{\rm Im}(y^3x-x^3 y)$ leading immediately to \impart.

Once the imaginary part of $r$ on any of the four cut edges has been found
the imaginary part on the other three edges can be determined by analytic
continuation. Then we can write down a dispersion relation for the
function $r(x)$ to fix its values on the real axis.
\ifig\contours{The cut structure of the eigenvalue density in the
critical region and the contour $C$ used in the dispersion relation. 
The same cuts and integration contour will arise in our analysis of the 
Kosterlitz--Thouless phase transition in section 4.}
{\epsfysize 2.5in\epsfbox{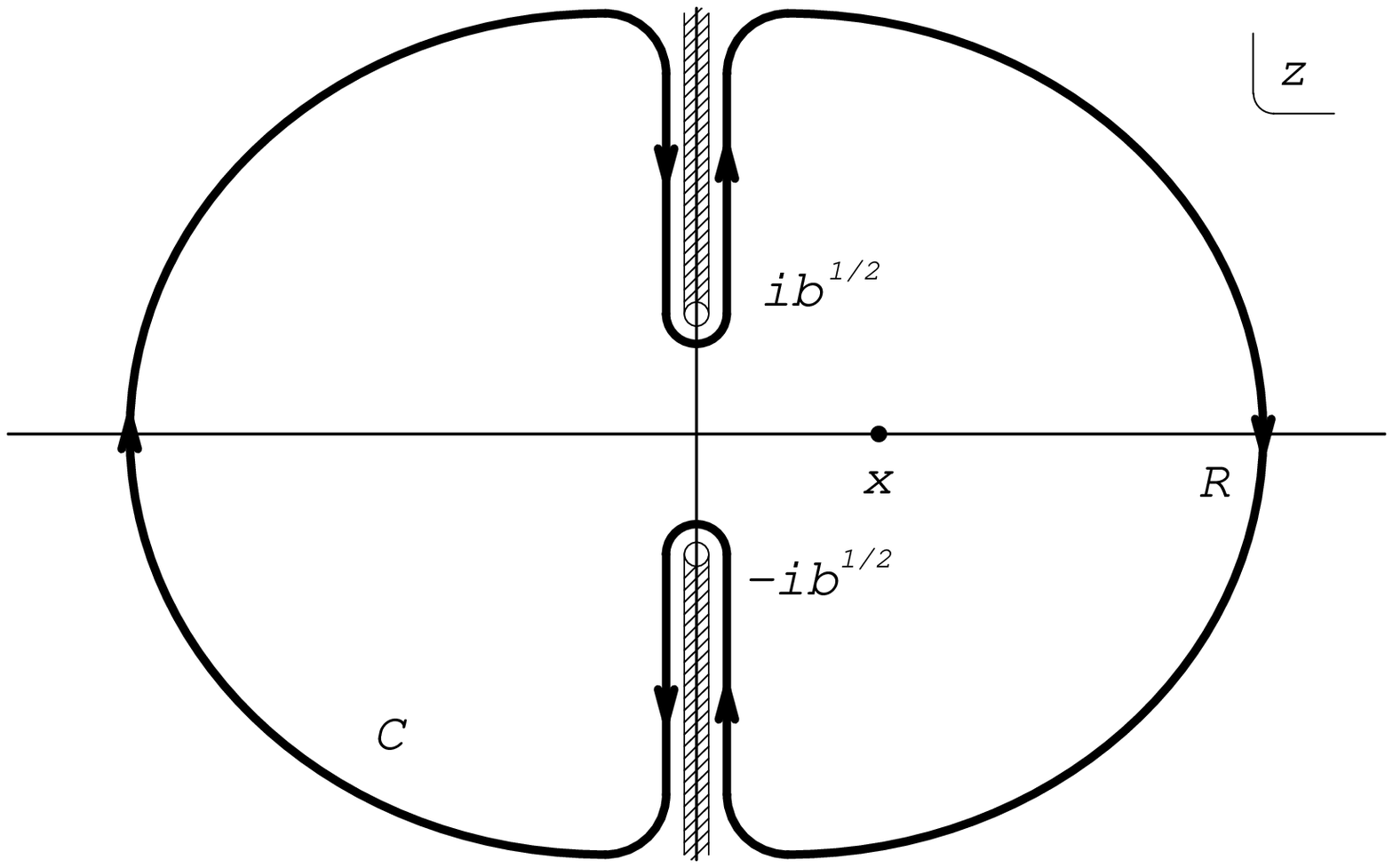}}

To this effect consider the integral
\eqn\intI{
I(x)=\oint {r(z)\, dz\over z^5 (x-z)}}
taken along the contour $C$ shown in the figure. The large radius loops 
closing $C$ at infinity do not make any contribution to this integral.
This follows from the nature of our approximation. Indeed, let us choose
the loop radius $R$ to be much larger than $b$ but still much smaller than 
the characteristic size of the potential $V(M)$. For $|z|<R$ the function 
$r(z)$ depends only on $z$ and $b$ and has the general scaling form $r(z)=
b^2s(z^2/b)$. However, for $z^2\gg b$ the leading asymptotic behavior of 
$r(z)$ should be independent of $b$, simply because $r(z)$ has a finite
limit when $b\to 0$, given by \rhosollog. 
Thus for large $\xi =z^2/b$ the function $s$ behaves
as $s(\xi)\sim \xi^2$ so that $r(z)\sim z^4$, perhaps with logarithmic
corrections. As a consequence for large $z$ the integrand in \intI\
decays as $1/z^2$ and the contribution from infinity equals zero.

Therefore, the only parts of the contour that contribute to $I(x)$ 
are the four cut edges. In fact, due to the square root nature of the branch
points the real part of $r$ on the cuts cancels out of the integral 
so that 
\eqn\jump{\eqalign{
I(x)&=\int\limits_b^{+\infty}\Biggl\{
{i\delta r(iu)\over (iu)^5 (x-iu)}+{i\delta r(-iu)\over (-iu)^5 (x+iu)}
\Biggr\}\, du \cr
&=-4ix\int\limits_b^{+\infty}{du\over u^4}{(2u^2-b)\sqrt{u^2-b}
\over x^2+u^2}\cr}}
where 
$$\delta r(iu)=-\delta r(-iu)=-2iu (2u^2-b)\sqrt{u^2-b}$$
is the jump of the function $r(z)$ across the cut.

Alternatively, $I(x)$ can be evaluated by taking residues at $z=0$ and $z=x$.
This yields 
\eqn\resid{
I(x)=2\pi i \, {r(x)\over x^5}- 2\pi i \biggl({{\tilde r_0}\over x^5}+
{{\tilde r_1}\over x^3}+
{{\tilde r_2}\over x}\biggr)}
the numbers ${\tilde r_0}, {\tilde r_1}$ and ${\tilde r_2}$ 
being defined by $r(x)= {\tilde r_0} +
{\tilde r_1} x^2+ {\tilde r_2}x^4+{\cal O}(x^6)$.
This allows us to compute $r(x)$. Evaluating the integral in \jump\ and 
comparing the result to \resid\ we easily deduce
\eqn\rresult{
r(x)=r_0 +r_1 x^2+ r_2 x^4 -{x\over \pi}(2x^2 +b)\sqrt{x^2+b}\log\Biggl[
{\sqrt{x^2+b}+x\over\sqrt{x^2+b}-x}\Biggr].}
The new constants $r_0, r_1$ and $r_2$ which appear in \rresult\ are related 
to the residues ${\tilde r_0}, {\tilde r_1}$ and ${\tilde r_2}$
according to the formulas
$$\left\{\eqalign{
r_0=&{\tilde r_0}\cr
r_1=&{\tilde r_1}+2b/\pi\cr
r_2=&{\tilde r_2}+14/3\pi.\cr}\right.$$
At this point it is necessary to check that the expression we obtained 
satisfies the full functional equation \linfunpole. The reason is that
in the above analysis we threw out the real part of the functional 
equation taking  only its imaginary part. It turns out that the full 
functional equation is obeyed if we impose an additional constraint
$r_1=r_2 b$. Thus we finally get
\eqn\rresultagain{
r(x)=r_0 + r_2 x^2 (x^2+b)-{x\over \pi}(2x^2 +b)\sqrt{x^2+b}\log\Biggl[
{\sqrt{x^2+b}+x\over\sqrt{x^2+b}-x}\Biggr].}
The parameter $r_0=r(0)$ can be set to zero by adequately redefining $b$.
The parameter $r_2$ plays the role analogous to ${\tilde c}_2$ in 
\rhosollog\ and for the same reason cannot be determined by expansion methods.
Note though that this parameter controls a totally regular, analytic 
contribution. The nonanalytic part of $r(x)$ is, on the contrary, fully
unambiguous. 

If we take $b$ to zero we recover the logarithmic singularity 
in $\rho(x)$ already found in \rhosollog.
Indeed, given $r_0=0$ the small $b$ limit of \rresultagain\ becomes
$$r(x)=r_0 + r_2 x^4 -{2x^4\over \pi}\log \biggl({4x^2\over b}\biggr)$$
which is fully consistent with \rhosollog\ provided 
$r_2=-(2/\pi)\log(\lambda^2 b/4)$.

The expressions for the eigenvalue density we found in  this section can be 
used, in principle, to determine the critical behavior of the free energy.
We shall pursue this problem elsewhere.
Furthermore, our 
expressions indicate that the analytic structure of the full 
exact solution for the matrix chain is rather complex. When viewed as an
analytic function of the complex variable $\epsilon$ this solution would have
singularities at all points of the form $\epsilon_{pq}=\sin(\pi p/2q)$ (and, 
as a consequence, at all other $\epsilon\in[0, 1]$.)
Remarkably, certain elliptic functions, like the Dedekind function
$${\cal D}^{-1}(q)=q^{-1/24} \prod_{n=1}^{\infty}{(1-q^n)^{-1}}$$
with $q=\exp(4i\arcsin\epsilon)$ have an analogous singularity structure. 
It would be quite interesting indeed if 
the exact $\rho(x)$ for the matrix chain was actually connected to such
elliptic functions.

\newsec{Kosterlitz--Thouless Phase Transition in Infinite Matrix Chain}

Mathematically, the Kosterlitz--Thouless phase transition 
in the matrix chain is quite similar to the logarithmic singularities we 
have discussed in the previous section. This transition occurs at the point 
$\epsilon=1$ where the small $x$ expansion for the eigenvalue density
given by \rhoansw\ or \rhocorrfirst\ ceases to make sense.

In this section we shall demonstrate that the critical behavior of the 
eigenvalue density at $\epsilon=1$ is given by a logarithmic law
\eqn\llaw{
\rho(x)\propto {|x|\over \log\big[1/(\lambda|x|)\big].}}
Such behavior is somewhat atypical. Indeed, in matrix models
the leading 
singularity in $\rho(x)$ is usually powerlike, $\rho(x)\propto|x|^{\delta}$.
However, it is possible to prove that for $\epsilon=1$ a power
law would be 
inconsistent with the saddle point equation. Although the explicit proof
shall not be given here one can easily reconstruct it using the methods of
this section.

We can acquire some idea of what $\rho(x)$ should be by inspecting 
the $\epsilon<1$ solution \rhoansw. As $\epsilon\to 1$ the slope $\alpha$ in 
$\rho(x)\propto\alpha |x|$ tends to zero. This suggests that at 
$\epsilon=1$ and $x\to 0$ 
the density $\rho(x)$ should vanish faster than $\alpha |x|$ with any $\alpha$.
On the other hand, when $\epsilon=1$
the asymptotics $\rho(x)\propto |x|^{\delta}$ with
$\delta>1$ contradicts Euler equations. Therefore, it is natural to look for
$\rho(x)$ which vanishes faster than $|x|$ but slower than $|x|^{\delta}$
for any $\delta>1$.

Expression \llaw\ has exactly this property and, most importantly, 
is consistent with Euler equations.
The most convenient way to see this is to use the functional equation \gpmeq.
At $\epsilon=1$ the functions $G_+(x)$ and $G_-(x)$ are
\eqn\gpmepsone{
G_{\pm}(x)=-x +{gx^3\over 2} \pm i\pi\rho(x).}
Such form for $G_{\pm}$ is very special. Indeed, 
if  we neglect both the $gx^3/2$ and $i\pi\rho(x)$ terms altogether
the resulting functions
$G_{\pm}(x)=-x$ would nonetheless satisfy $G_+[G_-(x)]=x.$
From this point of view, the Kosterlitz--Thouless phase transition is just 
the most senior of the logarithmic singularities found in section~3.
At these singularities, which occur for $\epsilon=\sin(\pi p/2q)$, 
the small $x$ behavior of $G_+$ and $G_-$  is given by \gggg, or
\eqn\gcompositfixedpoints{
G_{\pm}(x)={\rm e}^{\pm i\pi p/q}x +{\cal O}
\biggl[x^3\log\biggl({1\over \lambda x}\biggr)\biggr].}
If we raise $G_+(x)$ or $G_-(x)$ to the functional power $2q$---that is,
compose $G_+(x)$ with itself $2q$ times, we shall get
$$\underbrace{G_+\Bigl[G_+\big[\dots G_+(x)\big]\dots\Big]}_{2q\ {\rm times}}
=x+\dots.$$
One could say that, up to the higher order terms, the functions $G_{\pm}$
are the fixed points of order $2q$ with respect to the operation of 
functional composition.
All the logarithmic singularities of the matrix chain are in one-to-one
correspondence with such fixed points. In this language, the 
Kosterlitz--Thouless phase transition is represented by the fixed point
of the lowest---second---order. Consequently, there should be no surprise
that at $\epsilon=1$ we  also encounter the logarithmic type of
scaling behavior.

Let us now add $\pi\rho(x)=A|x|/\log[1/(\lambda|x|)]$ with an unknown 
coefficient $A$
and expand $G_+[G_-(x)]$ around $x=0$. In doing this expansion we shall use
again the prescription which directs us to throw away all absolute value signs.
The justification for that will be given later. In short, we write
$$G_{\pm}(x)=-x \mp {i A x\over \log(\lambda x)}+\dots.$$
The terms denoted by dots include the $gx^3/2$ term coming from the
interaction potential and the higher order corrections to the eigenvalue
density.
Generally, we expect these corrections to be down by powers of 
$1/|\log (\lambda x)|$ or $\log|\log(\lambda x) |/|\log (\lambda x)|$, 
like the higher order terms in 
the series \seriesuniv. Expanding the right hand side of the functional 
equation around $x=0$ yields 
\eqn\junk{
\eqalign{
G_+\big[G_-(x)\big]&=-G_-(x)-{iA G_-(x)\over \log\big[\lambda G_-(x)\big]}
+\dots \cr
&=x-{iA x\log(\lambda x)}+{iA x\log(-\lambda x)}\biggl[1-{iA\over
\log(\lambda x)}\biggr]+\dots.\cr}} 
We can now replace $\log(-\lambda x)=\log (\lambda x)-i\pi$ and expand
\junk\ in powers of $1/|\log (\lambda x)|$ to get 
\eqn\junkcontd{
G_+\big[G_-(x)\big]=x+ {A(A-\pi)x\over \log^2(\lambda x)} +{\cal O}\biggl(
{x\over \log^3\lambda x}\biggr).}
Therefore, 
the functional equation $G_+[G_-(x)]=x$ is indeed satisfied to 
the leading order
provided $A=\pi$, in complete agreement
with \llaw.

Note that in the last line  of \junk\ we kept the terms of order
$x/\log^2 (\lambda x)$. 
At first sight this may seem redundant and even incorrect.
After all, exactly such terms have been completely neglected in the original
expression for $\rho(x)$ given by \llaw.
Nonetheless, these additional terms change only the next, ${\cal O}[x/|\log^3
(\lambda x)|]$ order of expansion. They cancel out from the 
$x/\log^2 (\lambda x)$ part of the series very much like the leading term 
itself, $\rho(x)\propto x/|\log(\lambda x)|$ contributes only 
$x/\log^2 (\lambda x)$ and not $x/|\log(\lambda x)|$ to the right hand side of
\junkcontd.

At this point we would like to show how one constructs the solution of Euler 
equations, in terms of $v$ and $\rho$ or the Hopf function $f$, which 
corresponds to \llaw. There are three reasons why this is required.
First, it will provide us with a method to compute all logarithmic
 corrections to 
\llaw, prove their universality and sum the correction series in \seriesuniv\
all at once. Second, we shall be able to justify our way of dealing with 
absolute value signs in \junk. Third, functional equations like
\gpmeq\ can have spurious solutions which must be detected and eliminated.
An example of such spurious solution is given by a power series
\eqn\spur{
G_{\pm}(x) =-x +{gx^3\over 2}\pm i \sum_{k=1}^{\infty}a_k x^{2k}}
with real coefficients $a_k$. By expanding $G_+[G_-(x)]$
in powers of $x$ and equating the result to $x$ one can find all $a_k$
in a recursive fashion without encountering apparent 
contradictions. Nonetheless,
there is no solution of Euler equations which would give 
$$\pi\rho(x, t=0)=\pi\rho(x, t=1)=\sum_{k=1}^{\infty}a_k x^{2k}$$
while obeying the boundary condition $v(x, t=0)=-2x+gx^3/2$.
In other words, a solution of the functional equation \gpmeq\ does not
always yield a solution of the full hydrodynamic boundary problem
\euler,\bbc. The reason is, when viewed as an analytic function of the 
complex variable $x$ the eigenvalue density $\rho(x)$ may have several
different regular branches.
One of them takes real values for real $x$ and represents the actual
eigenvalue density of the infinite matrix chain. 
The power series expansions we construct refer, by definition, only to this 
branch.
However,
the functional equation \gpmeq\
involves the eigenvalue density computed at a complex point, 
$\rho[G_-(x)]$ which might take us to an entirely different branch of $\rho$.
The relationship between this other  branch and the ``proper'' branch 
expandable in a series can in reality be very complicated.

We can check whether a given solution of 
\gpmeq\ suffers from branch choice problems by computing $v$ and $\rho$
at $t=1/2$. A solution which is consistent with the complete set of 
hydrodynamic equations \euler,\bbc\ shall have $v(x, t=1/2)=0$ as required
by the time reflection symmetry \vhalf. Let us prove that this is indeed 
the case for the logarithmic eigenvalue density \llaw. To this end 
we shall use the implicit solution of the Hopf equation given by formula
\implicit. For $\epsilon=1$ the initial velocity equals
$v_0(x)=-2x +{\cal O}(x^3)$ so that the initial value of the Hopf function is 
\eqn\hfin{
f_0(x)=-2x + {i\pi x\over \log\big[1/(\lambda x)\big]}+\dots.}
Then the value of $f$ at $t=1/2$, denoted $f_{1/2}(x)$ can be determined 
from the equation 
\eqn\constrfhalf{
{2ix\over \pi}\log\biggl\{\lambda\biggl[x-{1\over 2} f_{1/2}(x)\biggr]
\biggr\}=x-{1\over 2} f_{1/2}(x).}
We are interested in solving this transcendental equation only in the limit
of small $x$. It is easy to check that the relevant solution is given by
\eqn\fhalf{
f_{1/2}(x)={4i\over \pi} x \log\biggl({1\over {\tilde \lambda} x}\biggr)+\dots}
where the dots stand for the subleading corrections of order ${\cal O}
[x\log|\log({\tilde \lambda}x)|]$.
Indeed, for such $f_{1/2}$
$$\eqalign{
\log\biggl\{\lambda\biggl[x-{1\over 2} f_{1/2}(x)\biggr]
\biggr\}&=\log(\lambda x)- {\pi i\over 2}+\log\biggl[
{2\over \pi}\log\biggl({1\over{\tilde \lambda}x}\biggr)-1\biggr]\cr
&=\log x -{\pi i\over 2} +{\cal O}\big[x\log|\log x|\big]\cr}$$
where we neglected the real valued terms which involve double logarithms 
as well as real constant terms.  The imaginary constants, on the other hand,
can be kept for the reason explained after equation \junkcontd.
In this approximation the left hand side of 
\constrfhalf\ equals
$${2ix\over \pi}\biggl(\log x -{\pi i\over 2}\biggr)=
x- {2ix\over \pi}\log\biggl({1\over x}\biggr)$$
which up to the terms of order ${\cal O}
[x\log|\log({\tilde \lambda}x)|]$ coincides with $x-f_{1/2}(x)/2$.

The expression for $f_{1/2}$ in \fhalf\ is purely imaginary.
Therefore, the fluid velocity at $t=1/2$, given by the real part of 
$f_{1/2}$, does vanish. Let us now examine the corresponding fluid density,
\eqn\rrhalf{
\pi\rho_{1/2}(x)={4x\over \pi}\log\biggl({1\over {\tilde \lambda} x}\biggr)+
\dots.}
Naively, we could expect that this expression receives logarithmic corrections,
generally of order ${\cal O}
[x\log|\log({\tilde \lambda}x)|]$. However, in a most remarkable way
all such corrections happen to cancel.
To avoid any misunderstanding, it would be false to say that the matrix chain 
eigenvalue density itself does not get logarithmic corrections.
It does. But the quantity $\rho_{1/2}(x)$ is still very simple, much
simpler than $\rho(x)$. In particular, the infinity of logarithmic terms 
in the double series \seriesuniv\ are fully summarized by equation \rrhalf.
That is to say, the only corrections to \rrhalf\ are the ${\cal O}(x^3)$ terms 
due to the quartic interaction in the matrix potential.

To prove this we shall turn our arguments around. Imagine starting 
at $t=1/2$ with a droplet of the density given by \rrhalf\ and zero velocity.
Let the droplet move according to Euler equations. Then compute the 
density and velocity of the droplet at $t=1$. 

Certainly, by time reflection we can always reverse the droplet motion to 
reconstruct the $0\le t\le 1/2$ 
part of the trajectory. As a result the boundary
condition $\rho(x, t=0)=\rho(x, t=1)$ will be satisfied automatically.
The only remaining boundary condition one will have to check is the equation
on velocity
\eqn\veloc{
v(x, t=0)=-v(x, t=1)=-2x +{gx^3\over 2}.}
We shall now prove that, neglecting ${\cal O}(x^3)$ terms,
the logarithmic density of \rrhalf\ generates precisely such a velocity.
The density $\rho(x)\equiv\rho(x, t=1)$ that we shall find in the course of 
this proof  will then be the matrix chain eigenvalue density at the 
Kosterlitz--Thouless point.

The velocity condition \veloc\ can be checked conveniently using the 
implicit Hopf solution \implicit. We start at $t=1/2$ with the Hopf function
\eqn\hopfhalf{
f_{1/2}(x)={4ix\over \pi}\log\biggl({1\over {\tilde \lambda}x}\biggr)+
{\cal O}(x^3)}
and we must finish at $t=1$ with $f$ being equal
\eqn\ffin{
f_1(x)\equiv f(x, t=1)=2x+i\pi\rho(x) +{\cal O}(x^3).}
Since the elapsing time interval is $\Delta t=1/2$
the implicit solution \implicit\ imposes the following relationship 
between $f_{1/2}$ and $f_1$
\eqn\imprel{
f_1(x)=f_{1/2}\bigl[x-\half f_1(x)\bigr].}
This is a complex valued equation which has both real and imaginary parts.
However, the only unknown in  $f_{1/2}$ and $f_1$ is one real valued
function $\rho(x)$. Therefore, the internal consistency of equation \imprel\
is not automatically guaranteed. Rather, such consistency is an indication of 
the correct choice for $\rho_{1/2}(x)$ that we made in \rrhalf.

For the purpose of computational convenience let us represent $\rho(x)$ in 
the form
\eqn\rhsigma{
\pi\rho(x)=\pi |x|\sigma(x) +{\cal O}(x^3).}
Note that again the $x<0$ and $x>0$ regions do not interact. Therefore,
we can safely replace $|x|\to x$ in \rhsigma\ so long as we apply our answers only to $x>0$.
Then taking  \hopfhalf\ and \ffin\ into account, 
equation \imprel\
translates into
$$2x+i\pi x\sigma(x)=-2 x\sigma(x)
\log\biggl[-{i\pi\over 2} {\tilde \lambda} x\sigma(x)\biggr].$$
Using $\log(-iy)=\log y -i\pi/2$ we see that the imaginary part of this 
equation is satisfied identically. The real part yields then an equation
on $\sigma(x)$,
$$2x= -2x\sigma(x)\log\biggl[{\pi\over 2} {\tilde \lambda} x\sigma(x)\biggr]
$$
or, equivalently,
\eqn\sigg{
{1\over \sigma(x)}+\log\sigma(x) =\log\biggl({1\over \lambda x}\biggr)}
with $\lambda=\pi {\tilde \lambda}/2$. For small $x$ the solution of \sigg\
is approximately 
$$\sigma(x)\propto{1\over \log\big[1/(\lambda x)\big]}$$
in complete agreement with \llaw.

A remarkable property of equation \sigg\ is its universality.
Indeed, the interaction potential enters $\rho(x)$ only through the
${\cal O}(x^3)$ terms which did not matter in our analysis. Therefore, the 
logarithmic terms summarized by $\sigma(x)$ will remain the same for all, 
not only quartic, matrix interactions.
Such universality may indicate that equation \sigg\ contains certain
information about the continuum limit of the theory. To understand what 
precisely this information is would be very interesting. 
A more specific problem that can hopefully be addressed using \sigg\ is to 
find the free energy at $\epsilon=1$.
Once we know the critical index of the free energy $\gamma_{\rm str}$ 
we can find the effective central charge at $\epsilon=1$ and determine 
what kind of a conformal field theory the matrix chain describes at that
point.  If the relation between $\gamma_{\rm str}$ and the eigenvalue
index $\delta$, 
$$\gamma_{\rm str}=1-\delta$$
holds in this case, formula \llaw\ indicates that at $\epsilon=1$ the
string susceptibility is most likely zero. This is characteristic of $c=1$
theories. However, the logarithmic corrections to the free energy
at $\epsilon=1$ would certainly be different from the logarithmic corrections
in the $c=1$ theory on a straight continuous line.
Therefore, these corrections deserve more investigation.
For example, equation 
\sigg\ may have an interpretation in terms of vortices which 
start forming a strongly interacting system precisely at the point 
$\epsilon=1$.

To compute the free energy carefully one must know the eigenvalue 
density in a certain interval of coupling deviations $\delta g=g-g_{\rm cr}$.
For small values of $\delta g$ the eigenvalue density can be found, as before,
using the method of dispersion relations. 

Again, we shall look for $\rho(x)$ which has two close square 
root branch points at $x\sim \pm i\sqrt{b}$ 
and is an even real function for real values of $x$.
We shall seek an approximation to $\rho$ valid for $x\sim\sqrt{b}\ll a$
where $a$ is the typical width of the quartic potential.
Moreover, since the density $\rho_{1/2}(x)$ seems to have a simpler form than
$\rho(x)$ itself we shall continue working directly with $\rho_{1/2}(x)$
even for $b\ne 0$.

The eigenvalue density with the desired properties can be found as a series
in powers of $\eta=\sqrt{b}/x$. 
That is, we go to the region $\sqrt{b}\ll x\ll a$ which is still within
the limits of our approximation. In this region $\eta\ll 1$ is a good 
expansion parameter. Consequently, we may look for $\rho_{1/2}(x)$ in terms
of a series in powers of $\eta$ or, equivalently, of  $b$.  
We shall now demonstrate that the relevant,
consistent with the hydrodynamic picture series is given by 
\eqn\serhalf{
\rho_{1/2}(x)=\rho_{1/2}(x)\big|_{b=0}+\sum_{k=1}^{\infty} 
a_k {b^k\over |x|^{2k-1}}.}
This ansatz for $\rho_{1/2}(x)$ is not arbitrary. A similar expansion arises,
for example, in the $\epsilon<1$ phase of the matrix chain. There
$\rho(x)\sim\sqrt{b+\alpha^2 x^2}$ can also be expanded in a series
$$\pi\rho(x)=\alpha |x|\sqrt{1+{b\over \alpha^2 x^2}}=\alpha |x|+
\sum_{k=1}^{\infty}\left({1/2\atop k}\right) {b^k\over (\alpha |x|)^{2k-1}}$$
identical in form to \serhalf.
Both of these expansions are explicitly singular at $x=0$ but 
the singularity disappears when the full series is summed.

Most interestingly, expansion \serhalf\ is consistent with the 
Hopf equation and 
the boundary conditions for any set of real coefficients $a_k$.
To see this we simply repeat the computation of equations 
\hopfhalf---\sigg. The function $f_1(x)=f(x, t=1)$ is still given by equation
\ffin\ while for $f_{1/2}(x)$ we now have
\eqn\fhalfnew{
f_{1/2}(x)= {4i\over \pi} x\log \biggl({1\over {\tilde \lambda}x}\biggr) +i\pi
\sum_{k=1}^{\infty}a_k {b^k\over x^{2k-1}}+{\cal O}(x^3).}
Imposing the implicit relation \imprel\ we find that again the 
imaginary part of
\imprel\ is satisfied identically while the real part yields an 
equation on $\rho(x)$,
\eqn\rhonew{
2x=-2\rho(x)\log\biggl[{\pi\over 2}{\tilde \lambda}\rho(x)\biggr]+
{\pi^2
\over 2}\rho(x) \sum_{k=1}^{\infty}(-)^k a_k \biggl[{4b\over  \pi^2 \rho^2(x)
}\biggr]^k.}
The coefficients $a_k$ have been left undetermined by the Hopf equation.
Rather, they are constrained by the analytic structure of $\rho_{1/2}
(x)$. Indeed, if $\rho_{1/2}(x)$ has two cuts shown in fig.3 then the real
part of $\rho_{1/2}$ on the cut edges follows directly from \serhalf.
For example, on the right edge of the upper cut, where $x=iu+0$
\eqn\cutt{\eqalign{
r(iu)\equiv {\rm Re}\big[\pi \rho_{1/2}(x)\big]&={\rm Re}\biggl[
{4x\over \pi}\log\biggl({1\over{\tilde \lambda}x}\biggr)\biggr]\cr
&=2u.\cr}}
Using the square root nature of the branch points we can then perform
an analytic continuation and find the real part of $\rho_{1/2}$ on the other
three edges.

The whole $\pi\rho_{1/2}(x)$ can now be restored through a subtracted 
dispersion relation as in \intI. The contour of integration is the same
contour $C$ in fig.3 while the dispersion integral itself is a little
different,
\eqn\intID{
J(x)=\oint\limits_C{\pi \rho_{1/2}(z)\over z\sqrt{z^2+ b}}{dz\over x-z}.}
The integrand in \intID\ has been constructed in such a way that the 
contribution from infinite loops closing the contour $C$ vanishes (see
\serhalf) while
the cuts contribute only through the real part of $\pi\rho_{1/2}(iu)$,
\eqn\IDcomputed{
\eqalign{
J(x)&=\int\limits_b^{+\infty}{2i\, r(iu)\over (iu)\big(i\sqrt{u^2-b}\big)}
\biggl({1\over x-iu}+{1\over x+iu}\biggr) \, du\cr
&=-8ix \int\limits_b^{+\infty}
{du\over (x^2+u^2) \sqrt{u^2-b}}.\cr}}
On the other hand, the integral in \intID\ can be evaluated via residues
to give 
\eqn\residuesID{
J(x)=-{2\pi i \over \sqrt{b} x} r_0 + 
{2\pi i\over x \sqrt{x^2+b}}\pi\rho_{1/2}(x)}
where $r_0=\pi\rho_{1/2}(x=0)$. Comparing this to \IDcomputed\ and
evaluating the integrals we find
\eqn\rhohalfresultser{
\pi\rho_{1/2}(x)=
{r_0\over \sqrt{b}}\sqrt{x^2+b}-{2x\over \pi}\log\Biggl[{\sqrt{x^2+b}+x
\over \sqrt{x^2+b}-x}\Biggr].}
The constant $r_0$, similarly to $r_2$ in \rresultagain\ cannot be 
determined by considering the region $x\sim\sqrt{b}\ll a$. To fix $r_0$
one would need to construct $\rho_{1/2}$ for all, not only small values of $x$.
The analytic structure of such globally constructed $\rho(x)$ would
include two other branch points at $x=\pm x_0\sim a$. Indeed, in matrix 
models with polynomial potentials the 
eigenvalue density is usually nonzero only within a finite interval
$[-x_0, x_0]$. The endpoints $x_0$ are, as a rule, square root
branch points. From the viewpoint of small $x$ expansions, the square
root character of the outer branch points is a nontrivial property of 
the eigenvalue distribution which is not automatically guaranteed.
Therefore, imposing it is likely to constrain the additional parameters 
such as $r_0$ in \rhohalfresultser\ or $r_2$ in \rresultagain.
After all, a very similar phenomenon has occured when we imposed the
two cut structure on $\rho_{1/2}(x)$.
There requiring two square root branch points at $x=\pm i \sqrt{b}$
fixed at once an infinite number of coefficients $a_k$ which remained
unconstrained by the Hopf equation.
%

To summarize, when the Hopf equation or the functional equation
$G_+[G_-(x)]=x$ are combined with reasonable assumptions about the
analytic structure of the eigenvalue density the solution becomes
determined uniquely. In the next section we shall explain, using an example,
how such global solutions can be constructed.

\newsec{Exactly Solvable Interaction Potentials}

By now we have seen that the Euler equations, the Hopf equation and the 
functional equation \gpmeq\ are all very useful in studying 
the singularities of $\rho(x)$. In this section we shall show how the same 
equations can be utilized to construct certain ``exactly solvable''
potentials $V(M)$ for which $\rho(x)$ is an elementary function.
We shall demonstrate that for these potentials the expansion methods of
sections~3~and~4 do reproduce the exact results.

The basic idea of such exact solutions is very simple. One chooses at 
will a midway eigenvalue density $\rho_{1/2}(x)$ making sure it
is properly normalized. Then the fluid is allowed to evolve under Euler
equations. Finally, at $t=1$ one reads off the velocity $v(x, t=1)$
and the corresponding eigenvalue density $\rho(x, t=1)\equiv\rho(x)$.
The result is an exact solution---given by $\rho(x)$ ---for the model where 
the potential can be found from \boundcond,\oppos\
\eqn\exactpot{
U^{\prime}(x)=2\big[x-v(x, t=1)\big].}
The hardest part of the problem is to choose $\rho_{1/2}(x)$ correctly,
so that the resulting $U(x)$ is physically reasonable.
Here we shall present a construction \REFhopf\ which produces a Penner-type
double-well potential rather similar to the pure quartic potential of
sections~3~and~4.

To this effect consider a special $\rho_{1/2}(x)$ defined by the following
quadratic equation
\eqn\quadratic{
k(x^2+r^2)^2+m^2(x^2+r^2)-2 (x^2-r^2) +b=0}
where $r\equiv\pi\rho_{1/2}(x)/2$. This equation looks, of course, 
very artificial, and it is. As it turns out, the potential $U(x)$
corresponding to such a choice of $\rho_{1/2}$ is particularly simple.
Later on, we shall  replace the left hand side of \quadratic\ by a more 
general polynomial to get  a solution for the $\epsilon>1$ phase of the
matrix chain.

The potential $U$ and the eigenvalue density $\rho$ that follow from
\quadratic\ can be easily found.
As in \hopfhalf---\imprel\ we use the implicit Hopf solution
\eqn\imprelagain{
f_1(x)=f_{1/2}\big[x-\half f_1(x)\big]}
with $f_{1/2}(x)=i\pi\rho_{1/2}(x)$ and 
\eqn\definfone{
f_1(x)=x-{1\over 2}U^{\prime}(x)+i\pi\rho(x).}
To determine $f_1$ we replace $x\to x^{\prime}=x-f_1(x)/2$ and 
$r(x)\to r(x^{\prime})=\pi \rho_{1/2}(x^{\prime})/2$ in equation \quadratic.
As a consequence of \imprelagain\
$$r(x^{\prime})=-{i\over 2}f_1(x)$$
which we can substitute into \quadratic\ to get a quadratic equation on 
$f_1$
\eqn\quadfone{
(1-kx^2)(x-f_1)^2- m^2 x (x-f_1) - b+x^2=0.}
Solving this quadratic equation we obtain the final value of the function $f$
\eqn\ffinalvalue{
f_1(x)=x - {m^2 x\over 2(1-kx^2)}+ {i\over 1-k x^2}\sqrt{
{m^4x^2\over 4}+ (b-x^2) (1-k x^2)}}
and, therefore
\eqn\finaldata{\left\{
\eqalign{
U^{\prime}(x)&= {m^2x\over 1-kx^2}\cr
\pi\rho(x)&= {1\over 1-k x^2}\sqrt{
{m^4x^2\over 4}+ (b-x^2) (1-k x^2)}.\cr}\right.}
The original definition of the matrix chain deals, however, with a rescaled
potential $V(M)$ related to $U(x)$ by \potentials. In our model
\eqn\vprime{
V^{\prime}(x)={\kappa^2 x+ G x^3\over 1- G\epsilon^2x^2/2}}
where we introduced the ``scaled'' mass $\kappa$ and the coupling
constant $G$
\eqn\relationsbetweencouplings{
\left\{
\eqalign{
m^2&=2+\kappa^2\epsilon^2\cr
k&=\half G\epsilon^3.
}\right.}
As always, by rescaling all matrices in the matrix chain 
partition function \chain\ we can make $\kappa^2=V^{\prime\prime}(0)$ 
anything we want.
Changing the value of $\kappa^2$ simply changes the scale of $\epsilon$
and shifts the position of all critical points.
We shall adhere to the convention of \definvofm\ that was used for 
pure quartic potentials and set $\kappa^2=-4$. In terms of the original
parameters $m^2$ and $k$ this would mean
\eqn\whatwoulditmean{
\left\{\eqalign{
m^2&=-2(2\epsilon^2-1)\cr
k&=\half G\epsilon^3,\cr}\right.}
a relation quite similar to \parameters.

The resulting potential $V(x)$ is plotted in fig.4. It goes to infinity as 
$x\to \pm x_{\rm max}=\sqrt{2/\epsilon^2G}$ thereby restricting the 
eigenvalues to a finite range $[-x_{\rm max}, x_{\rm max}]$. 
Note that the plot of fig.4 makes sense only when $\kappa^2 x_{\rm max}+
G x_{\rm max}^3>0$ ---otherwise for $x\to \pm x_{\rm max}$
the potential goes to minus, not plus infinity and therefore is 
unbounded from below. For $\kappa^2=-4$ this constraint translates into 
$\epsilon<1/\sqrt{2}$ or, equivalently, $m^2>0$.
\ifig\plots{An exactly solvable Penner-type potential $V(x)$ for
(right) $\kappa^2<0$ and $G>0$ or (left) $\kappa^2>0$ and $G<0$.}
{\epsfxsize2.0in\epsfbox{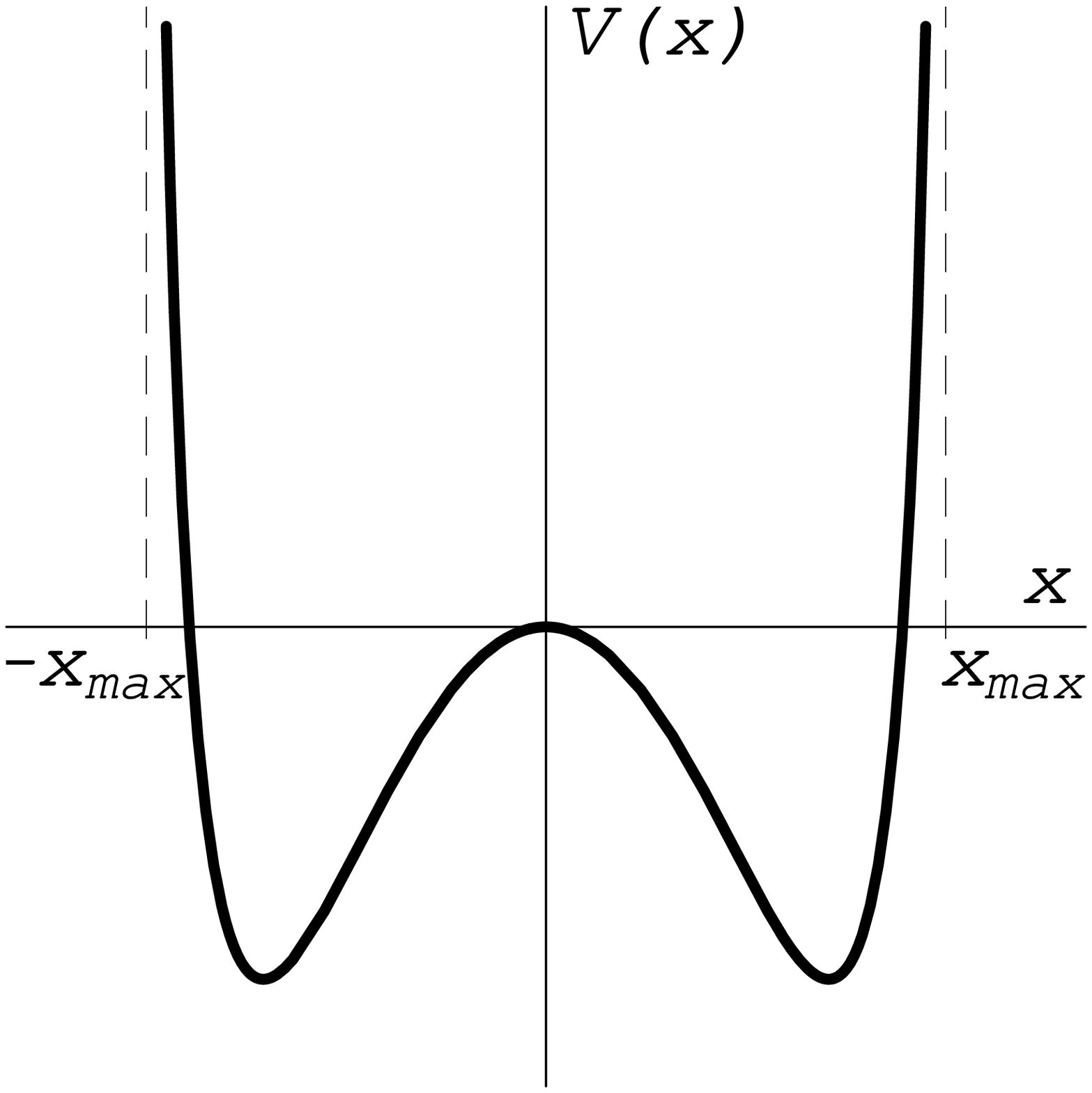}\hskip0.5in
\epsfxsize2.0in\epsfbox{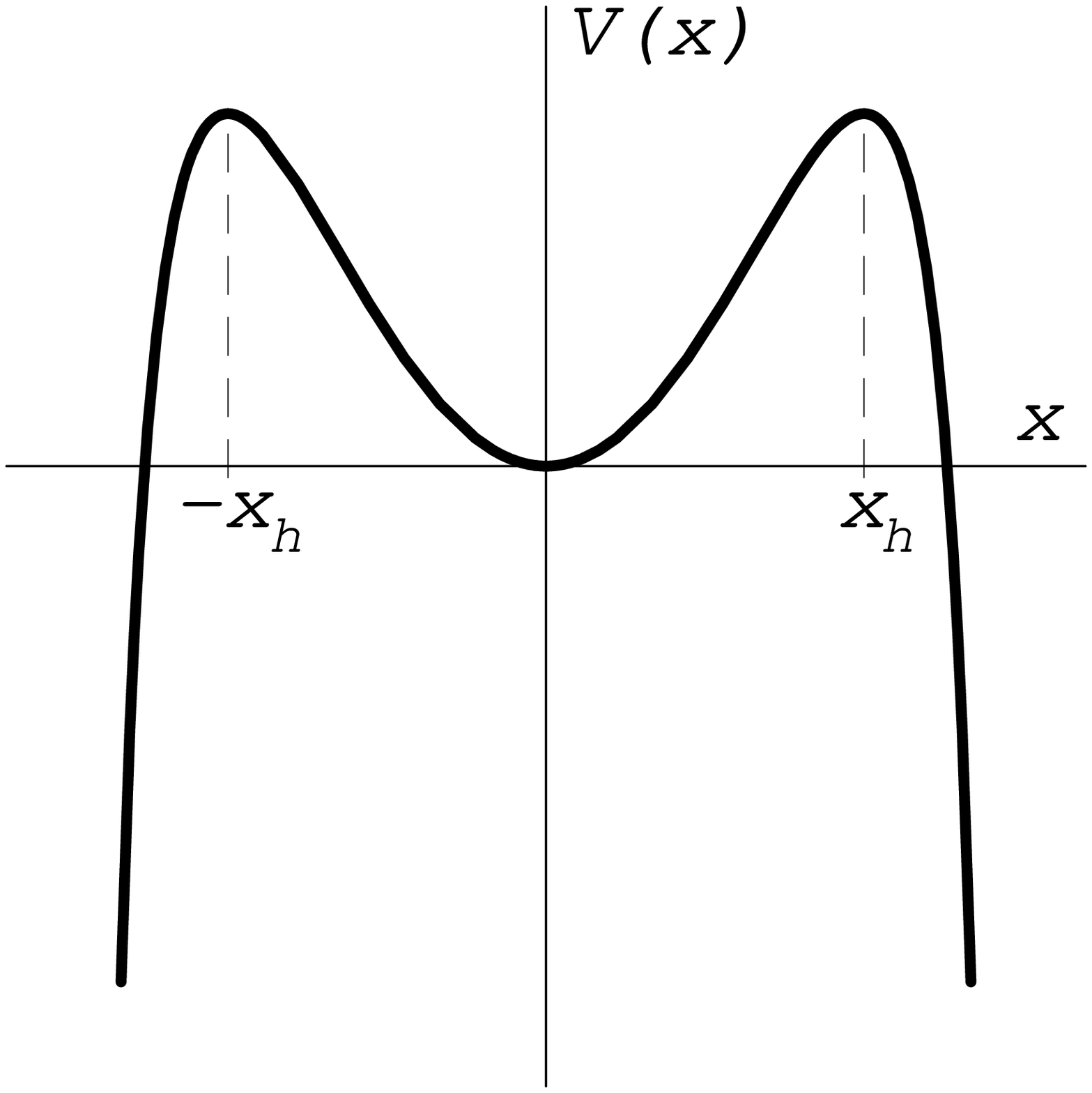}}
Such a picture is certainly consistent with what we already know. 
Indeed, at $\epsilon=1/\sqrt{2}$ we expect $\rho(x)$ to develop a
logarithmic singularity.
On the other hand, the solution \finaldata\ cannot, by construction,
have such logarithms. Therefore, it is perfectly natural that this 
solution should degenerate precisely at $\epsilon=1/\sqrt{2}$.
Furthermore, our exact solution has no singularities for $\epsilon$ 
between $0$ and $1/\sqrt{2}$. Here is an example of a special situation
when the terms responsible for such singularities---say, the terms in 
the curly brackets in \vanishsum--- are arranged to vanish. In fact,
the exact eigenvalue density \finaldata\ can be used to verify
and confirm all of the perturbative techniques developed in section~3.
For instance, when $b=0$ adapting the small $x$ asymptotic series of 
\rhocorrfirst\ to the case of our nonpolynomial potential
yields precisely the expansion of the exact solution \finaldata\ 
in powers of $x$. Moreover, for $b\ne 0$ it is possible to calculate,
up to arbitrarily high orders, the small $b$ expansion similar to 
\rhocorrbnotzero. The result, not surprisingly, can be summed into an
exact formula thereby providing a nontrivial check of our computational 
methods.

Quite remarkably, the same exact solution \finaldata\ can be used
to rediscover not only the singularity at $\epsilon=1/\sqrt{2}$
but also the Kosterlitz--Thouless phase transition. To see this
consider the same potential \vprime\ but with $\kappa^2>0$ and $G<0$.
Such a potential, also sketched in fig.~4, has two humps at $x_{\rm h}=\pm
\kappa/\sqrt{|G|}$ which play exactly 
the same role as the single hump at $x=0$.
By analogy with \definvofm\ we shall fix the
second derivative of the potential at the hump to be 
$V^{\prime\prime}(x_{\rm h})=-4$.
This imposes the following restriction on $\kappa$
\eqn\kappaofeps{
\kappa^2={2\over1-\epsilon^2}.}
That is, the potential degenerates as we approach the Kosterlitz--Thouless
phase transition at $\epsilon=1$. At that point expression \finaldata\
cannot possibly give the correct eigenvalue density. 
Such ``nontranscendental'' form of $\rho$ does not contain the requisite
logarithmic singularity which, as we found, has to be present for any,
even nonpolynomial, interaction potential.

Most likely, one could use the same methods to construct the 
exact solutions that would describe the logarithmic singularities themselves.
The idea would be to invent a globally defined $\rho_{1/2}(x)$ 
having both the small $x$ and small $b$ expansions consistent,
to first order, with \rhosollog, \rresultagain\ and \rhohalfresultser.
Then evolving it to $t=1$ would, by construction, generate the $\rho(x)$
with an adequate amount of logarithms. After the global $\rho(x)$ has been
found the direct computation of the free energy and various critical 
indices becomes reasonably straightforward. It would certainly be most 
interesting to actually carry out this program, especially for the
Kosterlitz--Thouless phase transition where the results would have a direct
meaning in terms of string theory.

\newsec{Interaction of Vortices with Curvature Defects}

We have seen throughout this paper that the Kosterlitz--Thouless
phase transition and the subleading logarithmic singularities 
of the $\epsilon<1$ phase bear remarkable mathematical similarity.
This similarity is perhaps expressed best by the formulas for $\rho_{1/2}(x)$.
Indeed, an explicit expression for $\rho_{1/2}$ at $\epsilon=1/\sqrt{2}$,
\eqn\rhohalfmiddle{
\pi\rho_{1/2}(x)=\sqrt{2b +4x^2 +g u(x)}}
with
\eqn\uofx{\eqalign{
u(x)=&
{8\over \pi}x\biggl(2x^2+{b\over2}\biggr)\sqrt{x^2+{b\over 2}}
\log\Biggl[{\sqrt{x^2 +b/2}+x\over \sqrt{x^2 +b/2}-x}\Biggr]\cr
&-2r_2\biggl(2x^2+{b\over2}\biggr)^2-b\biggl(2x^2+{b\over2}\biggr)\cr}}
has precisely the same structure as its analogue for $\epsilon=1$,
equation \rhohalfresultser.

Such parallels suggest that the qualitative physical picture of logarithmic
singularities must be related to the vortex picture of the 
Kosterlitz--Thouless transition. Below we shall argue that this is indeed the 
case.

In continuum theory, 
the action for a one-dimensional noncritical string compactified on a circle
of radius $R$ is
\eqn\action{
S[g, X]={1\over \pi}\int d^2 x \sqrt{g} \, g^{\alpha\beta}\partial_{\alpha}X
\partial_{\beta}X}
where $X$ is a compactified string coordinate, $x\equiv X+2\pi n R$ for any
$n\in\IZ$ and $g_{\alpha\beta}$ ---the two-dimensional worldsheet metric.
For flat $g_{\alpha\beta}$ a typical vortex---a solution of the
equations of motion
with a nonzero winding number $n$ ---is given by 
\eqn\vortex{
X(r, \phi)=n\phi R}
$r$ and $\phi$ being the polar coordinates parametrizing the flat area
of the worldsheet. The value of the action for such a vortex would be infinite
were it not for the cutoffs. Let us therefore introduce the infrared cutoff
$L$ and the ultraviolet cutoff $a$ (say, the lattice spacing if the worldsheet
is discretized.) With these cutoffs the vortex action equals
\eqn\vortaction{
S_{\rm vort}=2n^2 R^2  \log\biggl({L\over a}\biggr).}
The total weight with which a vortex contributes to the partition function
of the whole theory can be found by multiplying the number of places 
where the vortex could be centered (which is simply the total 
number of lattice sites $N_{\rm tot}\sim(L/a)^2$) 
by the Gibbs factor $\exp(-S_{\rm vort})$
\eqn\trans{
w_n\propto\biggl({L\over a}\biggr)^2 {\rm e}^{-S_{\rm vort}}=
\biggl({L\over a}\biggr)^{2(1-n^2R^2)}.}
As a result, depending on the value of $R$  vortices may or may not 
be important \REFkosterlitzthouless. 
If $R>1$ one has $2(1-n^2R^2)<0$ for all integer $n$ and
the total weight of any vortex vanishes
when the cutoffs are removed.  In this case  vortices can be neglected.
On the contrary, for $R<1$ the weight $w_1$ goes to infinity whenever 
$L/a\to\infty$ and so
vortices do matter. The two phases which arise 
are separated by a Kosterlitz--Thouless
phase transition with the critical value of compactification
radius $R_{\rm cr}=1$.

Let us now imagine that the two-dimensional worldsheet is not completely
flat but 
can have freely moving conical singularities. Around a conical
singularity such a worldsheet can still be described by the polar coordinates
$r$ and $\phi$ but with $\phi$ running in a different interval
$0\le\phi\le 2\pi +\alpha$. The ``excess angle'' $\alpha$ characterizes 
the amount of curvature acquired by the manifold due to the 
conical singularity.

For a vortex centered right at the top of the cone the vortex field $X$
would be given by 
\eqn\xcone{
X(r, \phi)={2\pi\over 2\pi+\alpha}n \phi R.}
Consequently, the vortex action changes as well,
\eqn\svortcone{
S_{\rm cone}={4\pi\over 2\pi+\alpha}n^2 R^2 
 \log\biggl({L\over a}\biggr).}
As we see, if $\alpha>0$ ---that is, when the conical singularity contributes
negative curvature---the value of the vortex action decreases compared to 
flat space.
Furthermore, in two-dimensional gravity
it does not cost any extra energy to create a conical singularity,
so long as the total worldsheet area and the genus remain the same.
Therefore, we could contemplate a Kosterlitz--Thouless phase transition
induced by such pairs of ``correlated'' cones and vortices centered closely
to each other. The critical value of $R$ or the corresponding 
$\epsilon=1/R$ for this transition can be 
found similarly to  \trans,
\eqn\critrad{
\epsilon_{\rm cr}(\alpha)=\sqrt{{2\pi\over 2\pi+\alpha}}<1.}
Of course, the pairs of cones and vortices with nearby centers 
form only a small susbset of all possible states in our system\foot{Here
is, perhaps, the most objectionable part of this argument. Strictly speaking,
the Kosterlitz--Thouless picture can be used reliably to infer the 
existence of a phase transition only for the whole system, and not 
for an artificially chosen
small 
subset of states. The only justification for what we say is that the results 
gotten via such reasoning do not contradict the exact computations of 
sections 3 and 4.}.
Roughly, the numbers of possible positions of either a vortex or
a conical tip are $(L/a)^2$ each, adding up to a total of $\sim(L/a)^4$ states.
Only $\sim(L/a)^2$ of them are the states where the cone and the vortex have
the same center. Consequently, the contribution of such ``correlated pairs''
can produce at most a finite size correction to the free  energy which
is down by $(a/L)^2\propto 1/A$ for a worldsheet of area $A$.
But this is exactly what we find from \rhosollog. There the logarithmic
correction to $\rho(x)$ is also subleading. Indeed, at $\epsilon=1/\sqrt{2}$
the leading contribution to the free energy---the one which is important 
in the double scaling limit---is associated with the first term of expansion 
for the eigenvalue density, $\rho(x)\propto |x|+\dots$. For any
$\epsilon<1$ this contribution is simply the standard $c=1$ free energy,
\eqn\freeen{
{\cal F}(\delta g)\propto {\delta g^2\over |\log\delta g|}+ 
{\cal O}(\delta g^3).}
The logarithmic correction $x^3\log[1/(\lambda x)]$ would modify only the 
${\cal O}(\delta g^3)$ terms of \freeen. And, as it turns out, such 
${\cal O}(\delta g^3)$ terms correspond precisely to $1/A$ corrections.

To see this, we just have to recall that the cosmological constant $\delta g$
and the worldsheet area $A$ are a pair of
thermodynamically conjugate variables.
That is, the string theory partition function with a nonzero $\delta g$,
given by ${\cal F}(\delta g)$, is related to the partition function 
of  random surfaces with fixed total area ${\cal Z}_{\rm str}(A)$
by the formula
$${\cal Z}_{\rm str}(\delta g)\equiv{\cal F}(\delta g)=\int\limits_0^{\infty}
dA\, {\rm e}^{-\delta g A} {\cal Z}_{\rm str}(A).$$
Consequently, an expansion in powers of $\delta g$ arising in \freeen\
would translate into a $1/A$ expansion for the fixed area partition function
and the fixed area free energy.
 
Finally, this picture can be used to explain why the logarithmic 
singularities form a discrete set. The reason is, for a discretized
surface generated via a matrix model the excess angle $\alpha$ at any graph
vertex can assume only certain discrete values. For example, a matrix model
with a cubic potential would produce surfaces composed of 
equilateral triangles. On such surface the angle $\alpha$ corresponding to
a vertex with the 
coordination number $q$ (where $q$ triangles meet together)
equals
$$\alpha={\pi\over 3}(q-6).$$
We see that $\alpha$ and therefore the critical lattice spacings 
computed from \critrad\ do come out discrete. However, this rough 
estimate fails to yield the exact critical values of $\epsilon$. 
This should not be a surprise. To get these critical values right 
one would have to take into account various fluctuations that we neglected.
For instance, the picture of an almost flat lattice with just a few 
curvature defects scattered around is certainly quite inaccurate.
In reality, almost every vertex of a generic two-dimensional graph
would have $q\ne 6$. It would  be quite interesting to find out 
whether such effects can be taken into account and whether our rough
picture survives that.

To summarize, the logarithmic singularities at $\epsilon<1$ carry along
lattice information. That is, their position and order depend on the 
particular type of polygons tiling our random surface. However, this 
does not yet  mean that such effects are of little interest. For example, the 
Kazakov's multicritical points in one-matrix models \REFkazakovmulticritical\
provide, also through
a very special way of tiling, a description of minimal models
coupled to quantum gravity. There the  connection with continuum theory
is not obvious from the 
matrix model formulation.  A similar indirect continuum interpretation
might exist  for the 
logarithmic singularities in the matrix chain.
In any case, the subleading 
singularities of the $\epsilon<1$
phase are quite an unusual phenomenon for matrix models. Indeed, 
in most theories the whole
singularity structure is captured by the double scaling limit of 
the matrix model, the corrections to this limit being perfectly regular.
For the matrix chain this does not seem to be so, which perhaps is rather 
satisfactory. As an example, it is very well known \REFgrossklebanov\ 
that the double scaling limit of the chain model is  smooth
throughout the interval $0\le\epsilon<1 $ and describes $c=1$ string theory.
This leads to a surprising conclusion that 
the discrete structure of the target space has no effect at all.
We see that although such a conclusion is absolutely true in the continuum
theory, the target space discreteness does show through a nontrivial 
set of corrections. 

Finally, the logarithmic singularities can be of interest 
for a separate reason. It appears 
remarkable that such a complicated family of singularities  arises 
precisely on the boundary between the $c<1$ and $c>1$ theories.
One might wonder, maybe somewhat groundlessly, whether a similar kind of 
structure could also emerge in any of the possible $c>1$ models.

\newsec{Conclusions}

We have investigated, in the large $N$ limit, the phase structure and the 
eigenvalue density of the one-dimensional infinite random matrix chain.
We developed a systematic expansion  applicable in the vicinity of the 
Kosterlitz--Thouless phase transition as well as throughout the $\epsilon<1$
phase of the model. Remarkably, such expansion can be used to extract the 
critical properties of the eigenvalue density even when the model is hard
to solve exactly. 

It turned out that the infinite matrix chain exhibits a few effects rather 
unusual in matrix models. In particular, we found an 
infinite number of subleading
(not important in the double scaling limit) critical points, similar in nature
to the Kosterlitz--Thouless phase transition. These special points appear to
arise from interactions of vortices with  
curvature defects and depend on the specific type of polygons tiling a
random surface.

The computational tools that were required here,
such as the hydrodynamic representation and the functional equation 
$G_+[G_-(x)]=x$ are likely to have many more applications. Even in the case
of the matrix chain these methods reproduce and generalize with great ease 
the results which would look nontrivial from other viewpoints.
They can also be used to explore the $\epsilon>1$ phase and to establish
the parallels with the character expansions.

In addition, there are a number of specific open problems that 
should be possible to resolve with our tools.
One such problem is
to compute carefully the free energy of the matrix chain at the
Kosterlitz--Thouless point. This could be done via  the 
procedure outlined at the end of section 5. Furthermore, it must be possible to
relate the critical behavior of the eigenvalue density $\rho(x)$ that we
now know, to the physical picture of vortex interactions.
Finally, it would be very interesting to evaluate, or at least to represent
in terms of the Hopf equation, the correlation functions of the matrix chain.
Such representation would help  to separate and compute the 
contributions of vortices with different vortex charges. As a 
consequence of this computation, we shall be able to study in a very direct 
fashion the
dynamics of vortices interacting with quantum gravity.

\bigskip
It is our pleasure to thank D. Gross, V. Kazakov and A.A. Migdal for their
interest, support and for helpful advice. We are especially
grateful to J. Goldstone
for a discussion concerning equation \linfunpole. Finally, we are indebted 
to E. Farhi and U.-J. Wiese for ideas regarding the 
numerical aspects of this problem.

This work is supported in part by funds provided by the U.S. Department of 
Energy under cooperative research agreement DE-FC02-94ER40818 and the
Swiss National Science Foundation.

\listrefs

\end